\documentclass{jfm}
\usepackage{graphicx}
\usepackage{epstopdf, epsfig}
\usepackage{amsmath,amsfonts}
\usepackage{caption}
\usepackage{subcaption}

\shorttitle{Liquid films falling down a vertical fiber}
\shortauthor{Y. Ruan, A. Nadim, L. Duvvoori, M. Chugunova}

\title{Liquid films falling down a vertical fiber: modeling, simulations and experiments}

\author{Y. Ruan\aff{1},
  A. Nadim\aff{1}, L. Duvvoori\aff{2},
 \and M. Chugunova\aff{1}}

\affiliation{\aff{1} Institute of Mathematical Sciences, Claremont Graduate University, CA 91711, USA  \aff{2} College of Engineering, University of California, Berkeley, CA 94720, USA}
\begin{document}

\maketitle

\begin{abstract}
We present a control-volume approach for deriving a simplified model for the gravity-driven flow of an axisymmetric liquid film along a vertical fiber. The model accounts for gravitational, viscous, inertial and surface tension effects and results in a pair of coupled one-dimensional nonlinear partial differential equations for the film profile and average downward velocity as functions of time and axial distance along the fiber. Two versions of the model are obtained, one assuming a plug-flow velocity profile and a constant thin boundary layer thickness to model the drag force on the fluid, the other approximating the drag using the fully-developed laminar velocity profile for a locally uniform film. A linear stability analysis shows both models to be unstable to long waves or short wavenumbers, with a specific wavenumber in that range having a maximal growth rate. Numerical simulations confirm this instability and lead to nonlinear periodic traveling wave solutions which can be thought of as chains of identical droplets falling down the fiber. Physical experiments are also carried out on such a system using safflower oil as the working liquid and a taut fishing line as the fiber. A machine learning scheme is used to find the best set of parameters in the laminar flow model to match the experimental results to the simulations. Good agreement is found between the two, with parameter values that are quite close to their original estimates based on the approximate values of the physical parameters.

\end{abstract}


\section{Introduction}
Liquid film flows along vertical cylindrical fibers exhibit complex and unstable interfacial dynamics with distinct regimes. Driven by the effects of Rayleigh-Plateau instability and gravity, a wide range of dynamics can be observed experimentally. These include the formation of discontinuous bead-like droplets, periodic traveling wave-like patterns, and irregularly coalescing droplets. The study of these dynamics has widespread applications in heat and mass exchangers, desalination \citep{sadeghpour2019water}, and particle capturing systems \citep{sadeghpour2017effects}, attracting much attention over the past two decades.

Depending on flow rate, liquid choice, fiber radius, and inlet geometry, three typical flow regimes have been observed \citep{kalliadasis1994drop,Ji2020Modeling}: (a) the convective instability regime, where bead coalescence happens repeatedly; (b) the traveling wave regime, where a steady train of beads flows down the fiber at a constant speed; and (c) the isolated droplet regime, where widely spaced large droplets are separated by small wave patterns. If other system parameters are fixed, and flow rate is varied from high to low, this can lead to flow regime transitions from (a) to (b), and eventually to (c).
Further analysis of the traveling wave patterns in regime (b) is expected to provide insights into many engineering applications that utilize steady trains of beads.

For small flow rates and thin films, classical lubrication theory is typically used to model the dynamics of axisymmetric flow on a cylinder. When the fluid film thickness is significantly smaller than the cylinder radius, \citet{frenkel1992nonlinear} proposed a weakly nonlinear thin-film equation to calculate the evolution of film thickness $h$ (or the height of the film) and capture both stabilizing and destabilizing effects of the surface tension in the dynamics. This evolution equation was further studied by \citet{kalliadasis1994drop}, \citet{chang1999mechanism}, and \citet{marzuola2019}. \citet{craster2006viscous} developed an asymptotic model which relaxes the thin film assumption, instead requiring that the film thickness be smaller than the capillary length. \citet{KDB} extended the thin film model to consider thick layers of viscous fluid by introducing fully nonlinear curvature terms. 
Recently \citet{Bert2019} investigated a family of full lubrication models that incorporate slip boundary conditions, fully nonlinear curvature terms, and a film stabilization mechanism. The film stabilization term, $\Pi(h) = -{A}/{h^3}$ with $A>0$, is added to the pressure and is motivated by the form of disjoining pressure widely used in lubrication equations \citep{reisfeld1992non} to describe the wetting behavior of a liquid on a solid substrate, and the scaling parameter $A > 0$ is typically selected based on a stable liquid layer in the coating film dynamics. Numerical investigations of experimental results in \citep{Bert2019} showed that compared to previous studies, the combined physical effects better describe the propagation speed and the stability transition of the moving droplets.

For higher flow rates where inertial effects are significant, coupled evolution equations of both the film thickness and local flow rate are developed \citep{trifonov1992steady, ruyer2008modelling, ruyer2009film, novbari2009energy}. These equations incorporate inertia effects and streamwise viscous diffusion based on the integral boundary-layer approach. Recently, \citet{Ji2020Modeling} further extended a weighted-residual integral boundary-layer model to incorporate the film stabilization mechanism to address the effects of the inlet nozzle geometry on the downstream flow dynamics. Finally, \citet{liu2021coating} have solved the full Navier-Stokes equations for film flow down a fiber directly using a domain mapping technique and have been able to reproduce the various flow regimes with remarkable accuracy. 

In this work we present a careful derivation of a simple new two-equation model, not starting from the Navier-Stokes equations, but based on a control volume analysis of the conservation of mass and momentum equations. While the approach is better justified if the axial velocity profile is a plug flow (i.e., uniform in the cross section as might be the case for a well-mixed turbulent flow), we can treat the drag force on the liquid film by the fiber wall in the laminar regime as well, obtaining a simple model that is suitable for viscous low Reynolds number flows. The next section provides the detailed derivation and is followed by the linear stability analysis of the system, showing that wavenumbers in a finite interval near zero are linearly unstable. Simulations of the full nonlinear equations with periodic boundary conditions in the axial direction show the emergence of finite amplitude steady traveling waves. We also carry out physical experiments on this system using a simple setup with safflower oil and fishing lines and capture images of the droplets that travel down the fiber. We show that our model can match the experimental results closely, with the best set of parameters obtained using machine learning, trained on a large set of simulation results with randomly chosen parameters near the physical range.  

\section{Model Derivation via Control Volume Analysis}

In this section we derive our model for an axisymmetric liquid film flowing down an infinitely long cylindrical fiber. Our approach is based on a control volume analysis of the conservation of mass and momentum equations, in which the axial velocity is replaced by a mean velocity that is uniform in the cross section but varies with axial distance and time. Assuming such a plug-flow profile greatly simplifies the derivation. However, one of the key terms that relates the viscous drag force on the fluid by the fiber is actually treated more carefully to make it consistent with the laminar flow profile for fully-developed flow down the fiber. Even if the flow is truly closer to a plug flow---e.g., in the high-Reynolds number turbulent regime where mixing causes the profile to be more uniform---we can still account for a drag force exerted between the solid surface of the fiber and the flowing film, proportional to the flow velocity, with some constant empirical coefficient related to a thin boundary layer thickness. As such, we end up with two versions of the model, one appropriate for low Reynolds number laminar flow and the other better suited to the high Reynolds number regimes. The models will appear quite similar though the scaling and the functional relation between the mean velocity and film thickness will be different between the two. Before deriving the model, it helps to compare and contrast these two cases in more detail, in the simpler situation when the flows are fully developed.

\subsection{Fully-Developed Flow}

\subsubsection{Plug Flow}

This case is simple to analyze. Consider a cylindrical fiber of radius $R$ and a liquid film whose interface is at distance $H$ from the fiber axis, resulting in a liquid film of thickness $H-R$. Suppose that the fluid is falling down the fiber under the influence of gravity at uniform speed $U$. At steady state (terminal draining velocity), the weight of any portion of the liquid between two axial locations is balanced by the drag force exerted by the solid surface of the fiber on the liquid. The weight of the liquid between two axial locations $x_1$ and $x_2$, with $\Delta x =x_2-x_1$,  is given by $\rho g \pi (H^2-R^2)\Delta x$. If the shear stress at the fiber surface is denoted by $\tau_{rx}$, the drag force exerted on that portion of liquid would be $2\pi R \tau_{rx}\Delta x$. Based on a dimensional reasoning, the form of the shear stress could be assumed to be
\[ \tau_{rx} =\frac{\mu U}{\ell}\,, \]
in which parameter $\ell$ is some quantity with units of length. It could be thought of as some measure of an extremely thin boundary layer thickness that might be separating the plug flow region with velocity $U$ from the fiber surface on which a no-slip boundary condition would exist. Of course, we ignore the boundary layer region when assuming plug flow, but still account for the drag force that the fiber exerts on the liquid. By balancing the weight of the liquid with the drag force, we can obtain a relationship between the flow speed $U$ and the film thickness $H$. The result is
\[ U = \frac{\rho g R \ell}{2 \mu} (h^2-1) \,,\]
in which $h=H/R$ is the ratio of liquid film radius to the fiber radius. If we assume parameter $\ell$ to be constant, the velocity scale can be chosen to be $U_o=\rho g R \ell /2\mu$ and the dimensionless draining velocity $u=U/U_o$ would be given by $u=f(h)=h^2-1$. We will compare this quadratic expression for the draining velocity as a function of $h$ with the result for fully developed viscous flow obtained below. We will find that this function $f(h)$ increases much more rapidly as $h$ increases away from 1, as compared to the situation with viscous laminar flow.

\subsubsection{Viscous Laminar Flow}

For the case of fully-developed laminar flow down the fiber, the velocity profile $u(r)$ can be obtained by integrating the axial component of the Navier-Stokes equation which reads
\[ \frac{\mu}{r}\frac{d}{dr}( r \frac{du}{dr}) + \rho g = 0 \,. \]
The boundary conditions are that $u(R)=0$ (no slip on the fiber surface) and $u'(H)=0$ (zero shear stress at the free surface). The resulting velocity profile is given by
\[ u(r) = \frac{\rho g R^2}{4 \mu} \left[ 1 - (\frac{r}{R})^2 + 2 (\frac{H}{R})^2 \ln(\frac{r}{R}) \right] \,. \]
The mean velocity $U$ can be calculated using the definition
$ U = {\int_R^H r u(r) dr}/ {\int_R^H rdr}$
resulting in
\[ U = \frac{2 \rho g R^2}{\mu} \frac{I(h)}{(h^2-1)} \]
with $h=H/R$ as before and 
\begin{equation}
\label{eq:Iofh}
I(h)=\frac{1}{16} \left(4 h^4 \ln (h)-3 h^4+4 h^2-1\right). 
\end{equation}
The shear stress at the fiber surface $\tau_{rx}=\mu u'(R)$ can be expressed as before in the form
\[ \tau_{rx} = \frac{\mu U}{\ell(h)} \]
but with length parameter $\ell$ now depending on $h$ and given by
\[ \frac{\ell(h)}{R} = \frac{4 I(h)}{(h^2-1)^2 }\,. \]
As such, the main difference between the plug flow model and the viscous laminar flow one is that in the former, $\ell$ is treated as a constant, whereas in the latter, it depends on the film thickness. With the proportionality constant between the shear stress and the mean velocity being dependent on $h$, the functional form of the dependence of the mean draining velocity on film thickness is quite different. In particular, the dimensionless mean velocity, now scaled with velocity scale $U_1 = 2\rho g R^2/\mu$, would be given by
\begin{equation}
\label{eq:laminarU}
    u(h) = \frac{U}{U_1} =\frac{I(h)}{h^2-1}=f_1(h) \,,
\end{equation} 
where function $I(h)$ is given by Eq.~(\ref{eq:Iofh}); this can be compared to the result for plug flow, which was $u(h)=f(h)=h^2-1$. The mean velocity $u$ varies much more as $h$ increases away from 1 for the plug flow model than for the laminar flow case. Close to $h=1$, $f(h)\approx 2(h-1)$ whereas $f_1(h)\approx (h-1)^2$. As such, for thin liquid films, the rate of increase of draining velocity with increasing film thickness is much stronger in the plug flow model than in the laminar flow one. 



\subsection{Control Volume Analysis}

\begin{figure}
    \centering
    \includegraphics[scale = 0.2]{"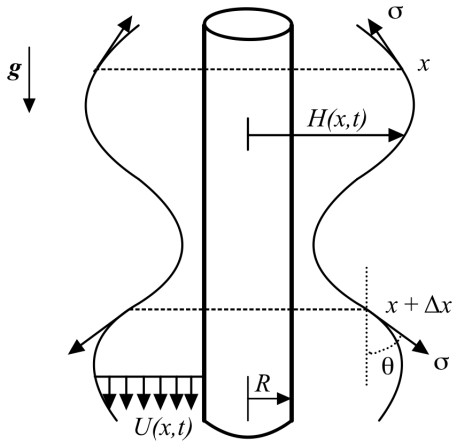"}
    \caption{ Schematic plot of a liquid film on a fiber. }
    \label{fig6:schematic}
\end{figure}

In order to derive the equations of motion for a falling film in which the film thickness varies with axial distance and time, i.e., $H=H(x,t)$, we use a control volume approach as depicted in Figure \ref{fig6:schematic}. We assume the velocity in the film to be uniform in the cross-section (interpreted as the mean velocity in the laminar case), but allow the latter to vary with axial location and time as well: $U=U(x,t)$. We consider a control volume consisting of the portion of the fluid between two axial locations $x$ and $x+\Delta x$, as shown in the figure. Denote the cross-sectional area of the fluid at any axial position and time $x$ by $A(x,t)=\pi (H^2(x,t)-R^2)$.

The integral form of the conservation of mass in the region between $x$ and $x+\Delta x$ reads
\[\frac{d}{dt} \int_x^{x+\Delta x} \rho A(x,t) dx = \rho AU|_{x} - \rho A U|_{x+\Delta x} \,,\]
equating the rate of change of mass to the rate at which mass enters the control volume at position $x$ minus the rate at which it leaves at position $x+\Delta x$. Based on the intermediate value theorem from calculus, the left-hand side of this equation can be written as
\[ \int_x^{x+\Delta x} \rho \frac{\partial A}{\partial t} (x,t)dx = \rho \frac{\partial A}{\partial t} (\xi,t) \Delta x \,,\]
where $\xi$ is somewhere in the interval $[x,x+\Delta x]$. Dividing both sides of the equation by $\Delta x$ and taking the limit $\Delta x \rightarrow 0$ results in the equation
\[ \frac{\partial A}{\partial t} + \frac{\partial (UA)}{\partial x} = 0 \]
for conservation of volume, as expected. Since $A(x,t)=\pi (H^2(x,t)-R^2)$, we can rewrite this equation as  
\begin{equation}
\label{eq:mass-dim}
2 H \frac{\partial H}{\partial t} +\frac{\partial (U(H^2-R^2))}{\partial x} = 0\,.
\end{equation}

Moving on to the conservation of linear momentum in the axial direction, one can similarly equate the rate of change of total linear momentum in the control volume to the net rate at which momentum flows into the control volume plus the sum of the forces in the axial direction acting on the fluid in that volume. This equation takes the form
\begin{align*}
\rho \Delta x \frac{\partial}{\partial t} (AU)|_\xi = & \rho (A U^2) |_x - \rho (A U^2) |_{x+\Delta x} + \rho g \Delta x A|_{\xi'}  \\
& +(pA)|_x - (pA)|_{x+\Delta x} + (A\tau_{xx})|_{x+\Delta x} - (A\tau_{xx})|_{x}\\
& - 2\pi R \Delta x \, \tau_{rx}|_{\xi''} +  2\pi \sigma (H \cos(\theta))|_{x+\Delta x} - 2\pi \sigma (H \cos(\theta))|_{x} \,.
\end{align*}
The terms on the right-hand side of this equation have the following physical interpretations: The first two terms provide the net rate at which momentum enters the control volume across the two boundaries, the next term is the weight of the volume of fluid in the control volume, the next two capture the contribution from the pressure force acting on the two cross-sections, followed by the two terms that account for any viscous normal stress at those same cross-sections, the next term is the drag force exerted on the fluid by the solid surface of the fiber, and finally, the last two terms capture the effect of surface tension acting on the perimeter of the free surface (since surface tension is tangent to the interface, to project it onto the axial direction, we need the cosine of the angle that the tangent vector makes with the axial direction in those terms). Points $\xi$, $\xi'$ and $\xi''$ are somewhere in the interval $[x,x+\Delta x]$; their precise location becomes irrelevant as $\Delta x$ tends to zero. Upon dividing this equation by $\Delta x$ and taking the limit $\Delta x \rightarrow 0$, we get the differential equation
\[ 
\rho \frac{\partial (AU)}{\partial t} + \rho \frac{\partial (AU^2)}{\partial x} = 
\rho g A -\frac{\partial (pA)}{\partial x}  + \frac{\partial (\tau_{xx} A)}{\partial x}
-2\pi R \tau_{rx} + 2\pi \sigma \frac{\partial(H\cos\theta)}{\partial x} \,. 
\]
Using the conservation of volume equation, the left-hand side of the last equation can be simplified to $\rho A (\partial U/\partial t + U\partial U/\partial x)$. Also, we substitute $\mu U/\ell$ for the shear stress $\tau_{rx}$ and $2\mu \partial U/\partial x$ for the normal viscous stress $\tau_{xx}$. Upon dividing the entire equation by the cross-sectional area $A(x,t)$ we thus obtain
\[ 
\rho (\frac{\partial U}{\partial t} + U\frac{\partial U}{\partial x}) +\frac{1}{A}\frac{\partial (pA)}{\partial x}= 
\rho g  + \frac{2\mu}{A}\frac{\partial}{\partial x}(A\frac{\partial U}{\partial x})
-\frac{2\pi \mu R U}{\ell A} + \frac{2\pi \sigma}{A} \frac{\partial(H\cos\theta)}{\partial x} \,. 
\]
In this equation, the cross-sectional area is given by $A(x,t)=\pi(H^2(x,t)-R^2)$, and since $\tan(\theta)=\partial H/\partial x$ (the slope of the free surface), the cosine of that angle is given by $\cos(\theta)=1/\sqrt{1+H_x^2}$ in which subscript refers to a partial derivative. The pressure within the film, $p(x,t)$, is taken to be uniform in the cross section and related by the Young-Laplace equation to the curvature of the free surface, namely $p(x,t)=\sigma \kappa(x,t)$, in which $\sigma$ is the surface tension and the curvature $\kappa$ is given in this geometry by
\[ \kappa(x,t)= \frac{(1 + H_x^2 - H H_{xx})}{H (1+H_x^2)^{3/2}}\,, \]
with subscripts referring to partial derivatives. Note that ordinarily the pressure in the fluid would be written as $p=p_o+\sigma \kappa$ in which $p_o$ is the constant pressure in the air outside the interface. However, in calculating the force on the control volume, the contribution of the force due to $p_o$ acting all around the control volume (including on the curved free surface) integrates to zero, so that constant part of the pressure is omitted.

The pressure term in the momentum equation can be written as a sum of two terms:
\[ \frac{1}{A}\frac{\partial (pA)}{\partial x} = \frac{\partial p}{\partial x} + \sigma \kappa \frac{1}{A} \frac{\partial A}{\partial x} \,. \] Interestingly, the second term on the right-hand side is exactly equal to the surface tension term on the right-hand side of the momentum equation, namely the term
\[ \frac{2\pi \sigma}{A} \frac{\partial(H\cos\theta)}{\partial x} \,, \]
so those two terms cancel each other leaving simply $\partial p/\partial x$ on the left-hand side of the momentum equation. The above cancellation is a consequence of a relationship that appears to be purely geometrical, involving the curvature $\kappa$ and the rates of change of area and the perimeter multiplied by the cosine factor,   namely: $\kappa \partial A / \partial x = 2 \pi \partial (H \cos\theta) / \partial x$ in which $\cos\theta=(1+H_x^2)^{-1/2}$. After this simplification, the momentum equation further divided by density $\rho$ becomes
\begin{equation}
\label{eq:mom-dim}
\frac{\partial U}{\partial t} + \frac{\partial}{\partial x} \left(\frac{1}{2}U^2 + \frac{\sigma \kappa}{\rho} \right) = 
g - \frac{2\pi \nu R U}{\ell A} + 
\frac{2\nu}{A}\frac{\partial}{\partial x}(A\frac{\partial U}{\partial x})\,. 
\end{equation}
Here $\nu=\mu/\rho$ is the kinematic viscosity of the fluid. Since $A=\pi(H^2-R^2)=\pi R^2(h^2-1)=\pi R^2 f(h)$,  upon choosing the velocity scale $U_o = g R \ell /2 \nu$ and defining the dimensionless velocity $u=U/U_0$, and upon scaling time with $U_o/g$, i.e., with $\hat{t}=g t / U_o$, the first term on the left-hand side and the first two terms on the right-hand side would yield a dimensionless equation of the form
\[ \frac{\partial u}{\partial \hat{t}} = 1 - \frac{u}{f(h)} \,. \]
Such an equation would hold if all $x$-derivatives were absent. It would suggest that for a given dimensionless film thickness $h$, the velocity $u$ of the film would relax exponentially in time to its terminal velocity $f(h)=h^2-1$ with a relaxation time of order one in dimensionless time $\hat{t}$.

Carrying the scaling further by nondimensionalizing the axial distance $x$ and curvature $\kappa$ with the fiber radius $R$ so that $\hat{x}=x/R$ and $\hat{\kappa}=R \kappa$, we obtain the fully nondimensional form of the axial momentum equation which, upon dropping the hats for clarity, reads
\begin{equation}
\label{eq:mom-nondim-plugflow}
u_t + (a u^2/2 + b\, \kappa)_x = 
[1 - u/f(h)] + c(h^2-1)^{-1}[(h^2-1)u_x]_x \,,
\end{equation} 
in which subscripts represent partial derivatives. The dimensionless curvature appearing in this equation is given by
\begin{equation}
\label{eq:curvature}
\kappa = 
\frac{1}{h(1+h_x^2)^{1/2}}-\frac{\partial}{\partial x}\left(\frac{h_x}{(1+h_x^2)^{1/2}}
\right) = \frac{(1 + h_x^2 - h h_{xx})}{h (1+h_x^2)^{3/2}}\,.
\end{equation}
Three dimensionless parameters, called $a$, $b$ and $c$, also appear in this equation, given respectively by:
\begin{equation}
\label{eq:parameters-plugflow}
a = \frac{U_o^2}{R g}= \frac{g R \ell^2}{4 \nu^2} \,, \qquad 
b = \frac{\sigma}{\rho R^2 g} \,, \qquad
c = \frac{2 \nu U_o}{R^2 g}=\frac{\ell}{R} \,.
\end{equation}
In its original form, parameter $a$ is seen to be the square of the Froude number and parameter $b$ is the reciprocal of the Bond or E\"{o}tv\"{o}s number. Parameter $c$ is the ratio of the characteristic boundary layer thickness to the fiber radius.

Using the same scaling, the dimensionless form of the conservation of volume equation takes the form:
\begin{equation}
\label{eq:mass-nondim}
   2 h h_t + a [u(h^2-1)]_x = 0 \,. 
\end{equation}

\subsection{Laminar flow case}

For the fully-developed laminar flow model, the above derivation proceeds similarly though in an approximate sense. The dimensionless mean velocity is now given by $u(h)=U/U_1=f_1(h)$ as shown in Eq.~(\ref{eq:laminarU}) with $U_1=2\rho g R^2/\mu$. This modifies the scaling and some of the dimensionless parameters in the model. Also, the balance of the gravity force and the radial derivatives in the viscous term produces the term $g[1-u/f_1(h)]$ on the right-hand side of the momentum equation where $f_1(h)$ is defined within Eq.~(\ref{eq:laminarU}). With those changes, and upon scaling time with $U_1/g$ and length with $R$, the dimensionless momentum equation takes a similar form to the one for the plug-flow model, i.e.,
\begin{equation}
\label{eq:mom-nondim-laminar}
    u_t + (a_1 u^2/2 + b \, \kappa)_x = [1 - u/f_1(h)] + c_1 (h^2-1)^{-1}[(h^2-1)u_x]_x \,,
\end{equation}
with parameter $b$ staying the same, while dimensionless parameters $a_1$ and $c_1$ differ from their earlier counterparts, now being given by
\begin{equation}
\label{eq:parameters-laminar}
    a_1=\frac{U_1^2}{Rg}=\frac{4 g R^3}{\nu^2} \,, \quad c_1=\frac{2 \nu U_1}{g R^2}=4 \,.
\end{equation}
The key assumption underlying this derivation is that the velocity in the axial direction can be replaced within most of the terms in the momentum equation by its mean over the liquid cross section. In particular, the square of the velocity on the left-hand side of the equation is replaced by the square of the mean velocity. Also, the pressure field is still assumed to be uniform in the cross section. Despite the approximate nature of this model, we shall see that it is successful in matching the experimental results fairly well with parameter values that are quite close to their theoretical values.

The dimensionless equation for conservation of mass (or volume) looks the same as Eq.~(\ref{eq:mass-nondim}) but with parameter $a$ replaced by $a_1$ in this case:
\begin{equation}
\label{eq:mass-nondim-laminar}
   2 h h_t + a_1 [u(h^2-1)]_x = 0 \,. 
\end{equation}

\subsection{Summary}

To summarize, our one-dimensional two-equation model for an axisymmetric liquid film falling down a vertical fiber consists of the equations for the conservation of mass (\ref{eq:mass-nondim}) and axial momentum (\ref{eq:mom-nondim-plugflow}) for plug flow, or the corresponding pair (\ref{eq:mass-nondim-laminar}) and (\ref{eq:mom-nondim-laminar}) for laminar flow. The dependent variables are the dimensionless film radius $h(x,t)$ and the mean axial velocity $u(x,t)$. The dimensionless film thickness is given by $h(x,t)-1$. The parameters for the plug flow case are given in Eq.~(\ref{eq:parameters-plugflow}) and those for laminar flow in Eq.~(\ref{eq:parameters-laminar}). Dimensionless curvature is given by the expression in Eq.~(\ref{eq:curvature}). We have used both of the forms given in that equation successfully in our numerical simulations. For the plug flow model, we have the function $f(h)=h^2-1$; for the laminar flow model, it is replaced by $f_1(h)=\left(4 h^4 \ln (h)-3 h^4+4 h^2-1\right)/\left(16(h^2-1)\right)$.
%
%
When performing numerical simulations, we choose a domain of dimensionless length $L$ in space, so that $x \in [0, L]$, and solve the equations up to a chosen final dimensionless time of $T$, so that $t \in [0, T]$. We apply periodic boundary conditions in $x$. Parameters $a,b,c$ (or $a_1,b,c_1$) are all positive and represent the effects of inertia, surface tension and axial viscous diffusion respectively.

\section{Linear Stability Analysis}
In this section, we conduct a linear stability analysis about constant solutions of the system (\ref{eq:mass-nondim})-(\ref{eq:mom-nondim-plugflow}) (plug flow) or (\ref{eq:mass-nondim-laminar})-(\ref{eq:mom-nondim-laminar}) (laminar flow). Note that any constant $h_0$ and $u_0$ that satisfy $u_0 = f(h_0)$ (or $u_0=f_1(h_0)$) is a solution of the system. We define the perturbed solution in the form below:
\begin{align*}
    h(x,t) &= h_0 + \epsilon h_1(x,t) + O(\epsilon^2) \\
    u(x,t) &= u_0 + \epsilon u_1(x,t) + O(\epsilon^2)
\end{align*}
where $h_0$ and $u_0$ are constants. Small parameter $\epsilon$ is introduced for bookkeeping purposes only. At $O(\epsilon)$ we derive a linearized system for the leading perturbations as
\begin{align*}
    \frac{\partial u_1 }{\partial t} + \frac{\partial}{\partial x} \left[ a u_0 u_1 - b ( \frac{h_1}{h_0^2} + \frac{\partial^2 h_1}{\partial x^2} )\right] &=  \frac{ f'(h_0) h_1}{ f(h_0)} - \frac{u_1}{ f(h_0)}  + c \frac{ \partial^2 u_1}{ \partial x^2} \\
    \frac{\partial h_1}{ \partial t} + \frac{\partial}{\partial x} \left[  u_0 h_1 + \frac{1}{2}u_1 (h_0 - \frac{1}{h_0}) \right] &= 0
\end{align*}
for the plug flow case. For laminar flow, $f(h_0)$ and $f'(h_0)$ should be replaced by $f_1(h_0)$ and $f'_1(h_0)$, and $a$ and $c$ are replaced by $a_1$ and $c_1$.

These being linear constant-coefficient equations, one can seek exponential solutions for $h_1$ and $u_1$ in the form:
\begin{align*}
    h_1 &= \Re\{\hat{H} e^{ikx + \alpha t}\} \\
    u_1 &= \Re\{\hat{U} e^{ikx + \alpha t}\}
\end{align*}
where $\hat{H}$ and $\hat{U}$ are complex amplitudes, $\Re$ denotes the real part, $k$ is the wavenumber (assumed real) and $\alpha$ is the growth rate. Upon substitution of these exponential forms into the linearized equations, we obtain the linear system
\begin{equation}
\label{eq:linsyst}
    \begin{bmatrix}
        i( 2 a k u_0 ) + 2\alpha & ika( h_0 - \frac{1}{h_0}) \\
        i( - \frac{kb}{h_0^2} + bk^3) - \frac{f'(h_0)}{f(h_0)}  & i(aku_0) + \alpha + ck^2 + \frac{1}{f(h_0)} 
    \end{bmatrix} \begin{bmatrix}
    \hat{H} \\
    \hat{U}
    \end{bmatrix} = 0\,.
\end{equation}
The above form is for the plug flow case; for laminar flow we replace $f$, $a$ and $c$ with $f_1$, $a_1$ and $c_1$.

\begin{figure}
\begin{subfigure}{.5\textwidth}
  \centering
  \includegraphics[width=.95\linewidth]{"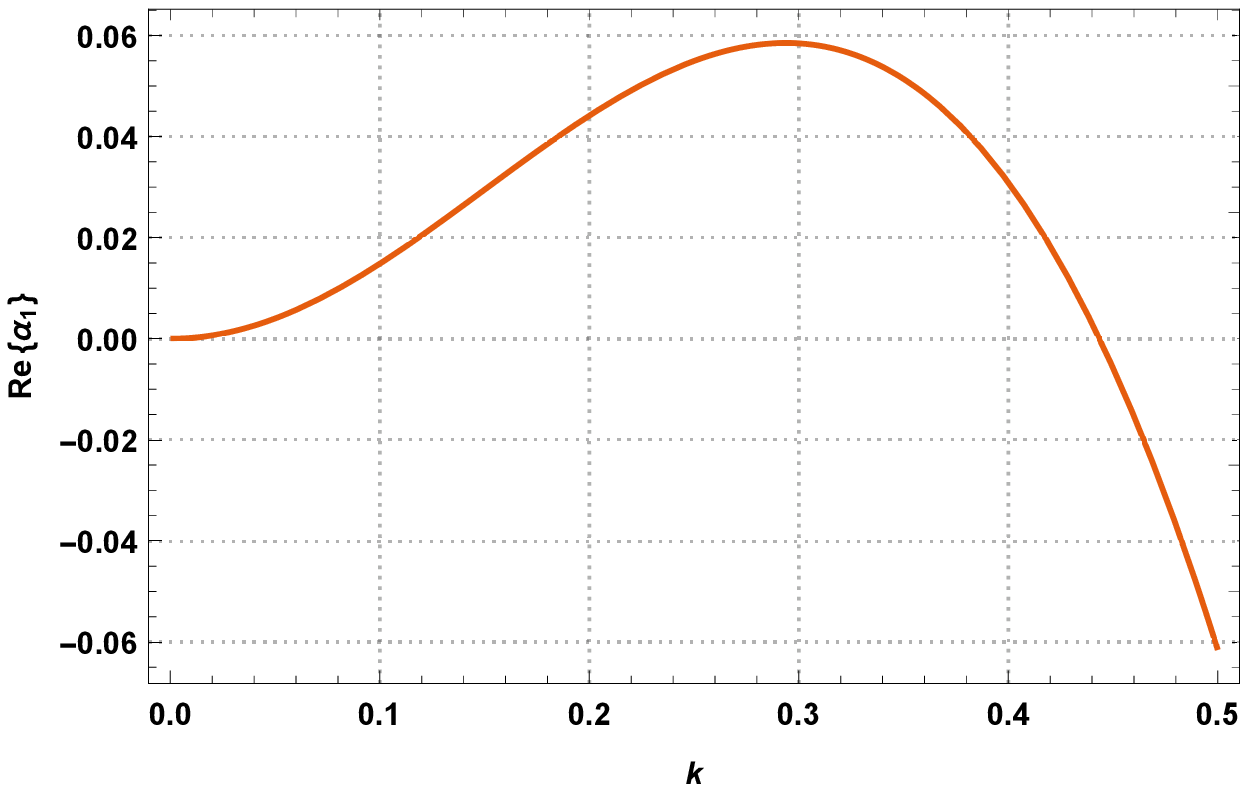"}
  \label{fig1:sub1}
\end{subfigure}%
\begin{subfigure}{.5\textwidth}
  \centering
  \includegraphics[width=.95\linewidth]{"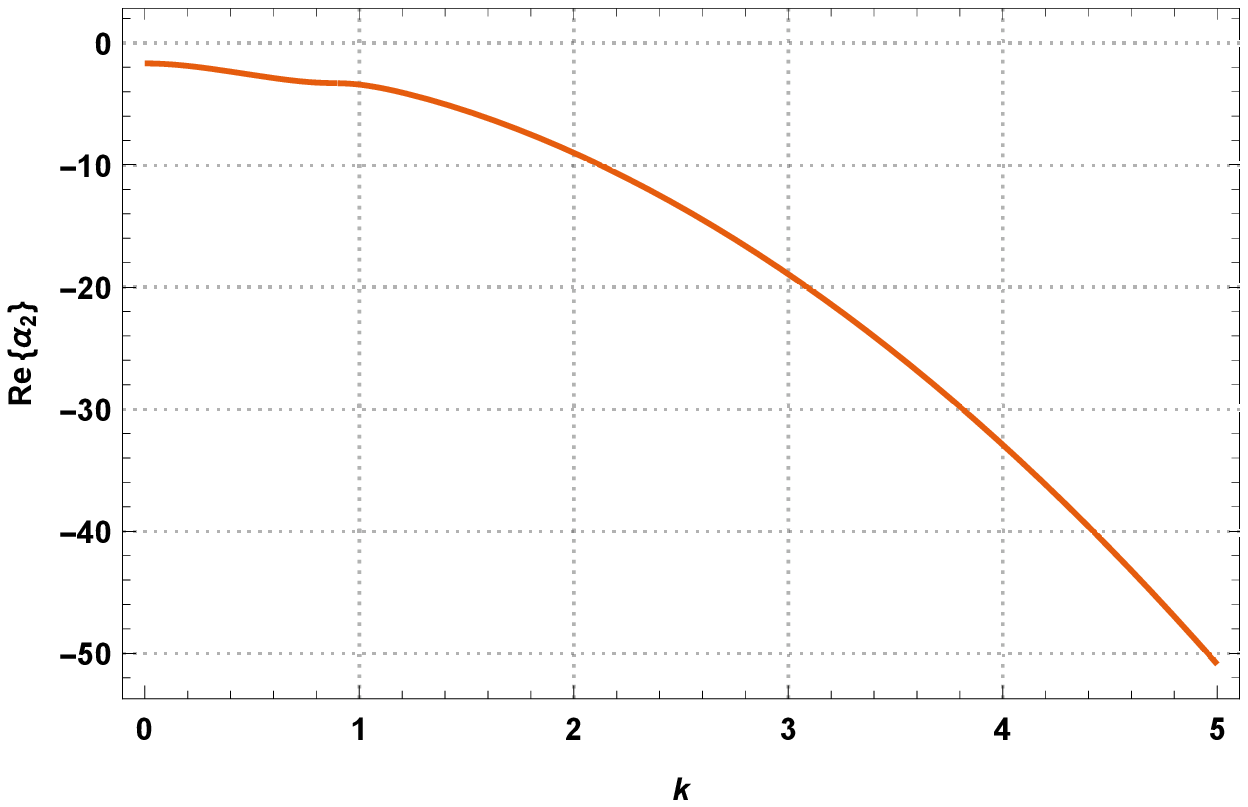"}
  \label{fig1:sub2}
\end{subfigure}
    \caption{Plots of $\Re(\alpha_1)$ and $\Re(\alpha_2)$ versus wavenumber $k$ for the laminar-flow model with $h_0 = 2.5$, $u_0 = f_1(h_0)$, $a_1 = 1$, and $b = 11$. While $\Re(\alpha_2)$ stays negative for all $k$, $\alpha_1$ has a positive real part over a finite range of wavenumbers $k$ near $k=0$, exhibiting a maximum growth rate at a wavenumber close to 0.3.}
    \label{fig1:stabilitylaminar}
\end{figure}

\begin{figure}
\begin{subfigure}{.5\textwidth}
  \centering
  \includegraphics[width=.95\linewidth]{"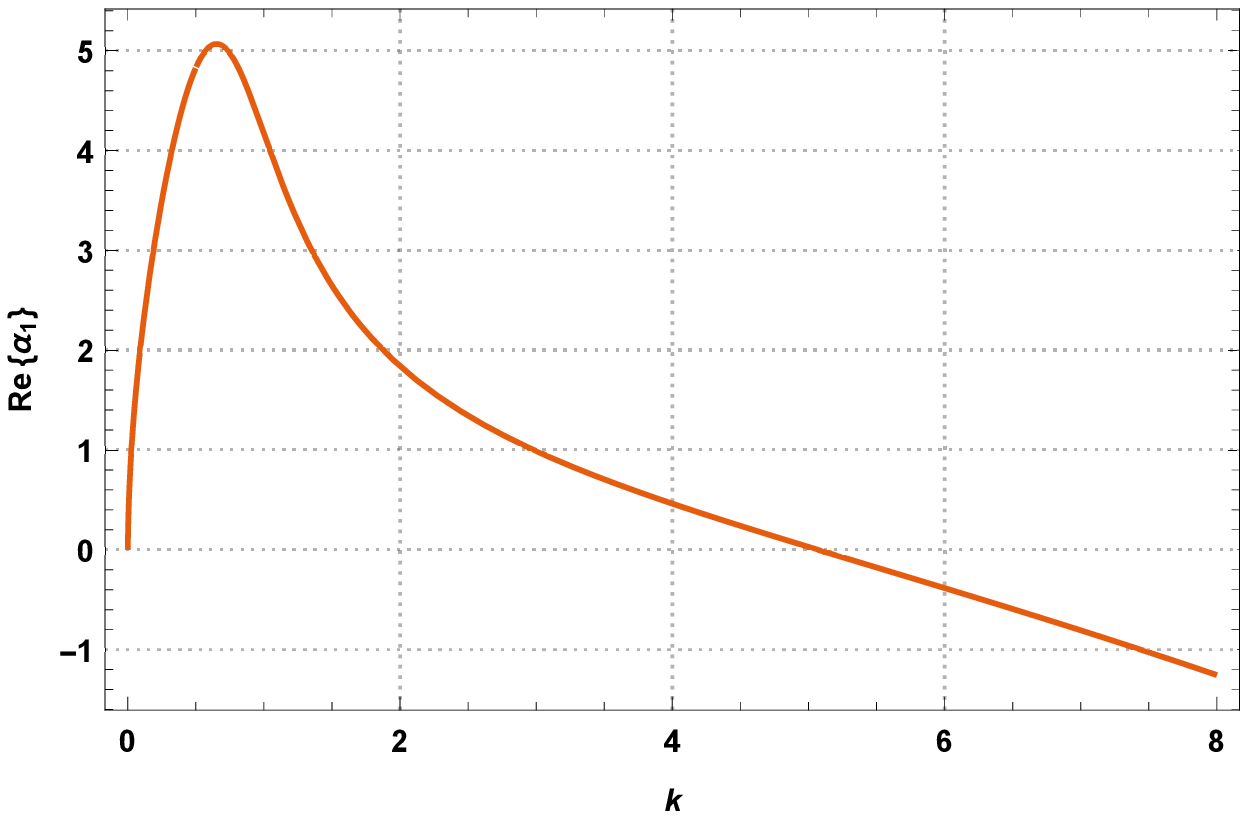"}
\end{subfigure}%
\begin{subfigure}{.5\textwidth}
  \centering
  \includegraphics[width=.95\linewidth]{"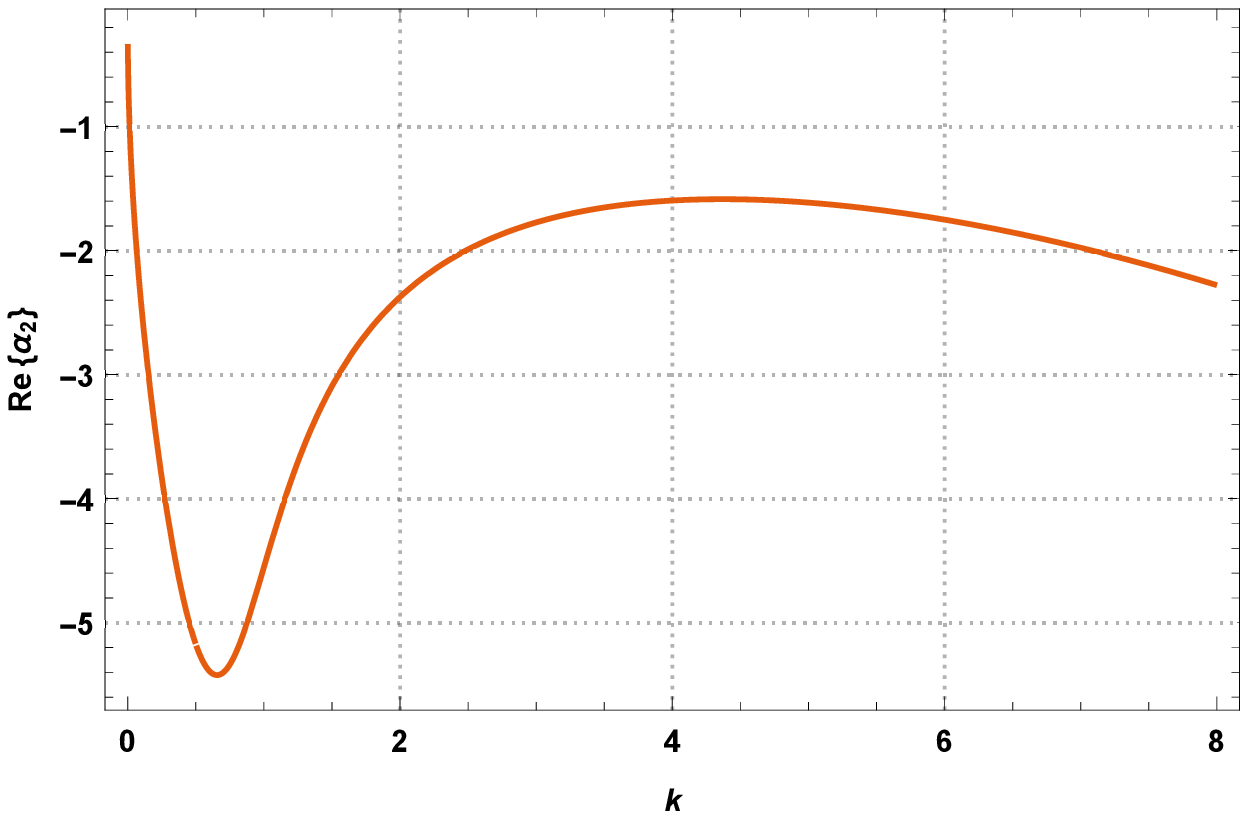"}
\end{subfigure}
    \caption{Plots of $\Re(\alpha_1)$ and $\Re(\alpha_2)$ versus wavenumber $k$ for the plug-flow model with $h_0 = 2$, $u_0 = f(h_0)$, $a = 100$, $b = 2$ and $c=0.05$. While $\Re(\alpha_2)$ stays negative for all $k$, $\alpha_1$ has a positive real part over a finite range of wavenumbers $k$ near $k=0$.}
    \label{fig1:stabilityplugflow}
\end{figure}

To have a non-trivial solution of the system (\ref{eq:linsyst}) for $\hat{H}$ and $\hat{U}$, the determinant of the coefficient matrix needs to be zero. This provides a quadratic equation for the growth rate $\alpha$, which can be solved analytically but involves lengthy expressions that are not displayed here. Given $h_0$, $u_0=f(h_0)$, and parameters $a$, $b$ and $c$, we obtain the two roots of the quadratic equation $\alpha_1$ and $\alpha_2$ (which are complex valued) as functions of the wavenumber $k$. The procedure is identical for the laminar flow model, although function $f_1(h)$ is involved and parameters $a_1$ and $c_1=4$. Figure \ref{fig1:stabilitylaminar} provides plots of the real parts of the two growth rates versus wavenumber $k$. These are for the laminar flow model with parameters: $h_0=2.5$, $a_1=1$ and $b=11$. These values are chosen since they are quite close to those for the experiments described in Sections~\ref{sec:exp} and \ref{sec:ML}. Note that if the real part of either growth rate is positive, waves of those wavenumber grow and the system is linearly unstable. In the figure, we see that one of the roots does indeed have a positive real part over a range of wavenumbers $k \in (0,k_{\tiny \mbox{max}})$, with a maximal growth rate occurring for some wavenumber in that interval. We thus see that uniform solutions are unstable to perturbations of small wavenumber or long wavelength. Figure~\ref{fig1:stabilityplugflow} is obtained for the plug-flow model with the set of parameters indicated in the caption, approximating the hypothetical high Reynolds number case discussed in Section~\ref{sec:disc}. We still find that one of the growth rates has a positive real part over a range of wavenumbers near zero. However, in this case the actual rate of growth is much higher. As a result, numerical simulations of the plug-flow model are more challenging since the function $f(h)$ varies a lot more than $f_1(h)$ and the growth rate of the instabilities is also quite a bit higher than for the laminar flow model. 

The stability result was checked against numerical simulations. We did a comparison between two simulations, one using the full nonlinear model and the other using the linearized one to see the effects of nonlinearity on the film thickness evolution. In the simulation, we took an initial profile $h(x,0) = h_0 + \epsilon \sin (kx)$ with a small $\epsilon$ and the domain length was chosen as $L = {2\pi}/{k}$. We chose the wavenumber $k$ as the most unstable wavenumber for $h_0$. We found that the rate of increase of the maximum film height does follow the simple exponential function (the logarithm of the perturbation growing linearly in time) for the linearized model, while for the full nonlinear model the growth slows down as time increases as a result of nonlinear interactions. 

\section{Simulations}

Using COMSOL Multiphysics we carried out many simulations of both the plug flow system (\ref{eq:mass-nondim})-(\ref{eq:mom-nondim-plugflow}) and the laminar flow one (\ref{eq:mass-nondim-laminar})-(\ref{eq:mom-nondim-laminar}) for various sets of parameters. For the results reported in this section, we took the domain $x\in[0,L]$ with $L=20$ and assumed periodic boundary conditions in $x$. For the initial condition, we took $h(x,0)=h_0+0.1 \sin(2 \pi x/L)$ for the film profile and $u(x,0)=f(h(x,0))$ or $u(x,0)=f_1(h(x,0))$ for the initial velocity. We integrated the equations to a final time ranging from several hundreds to several thousands until a steady traveling wave profile was obtained. We found that for smaller values of parameter $a$ it takes longer to reach a steady traveling wave shape. Focusing on the shape of the traveling wave profile, we took the final steady shape and centered its peak in the middle of the interval in order to be able to make the following comparison plots.

\begin{figure}
\includegraphics[height=1.6 in]{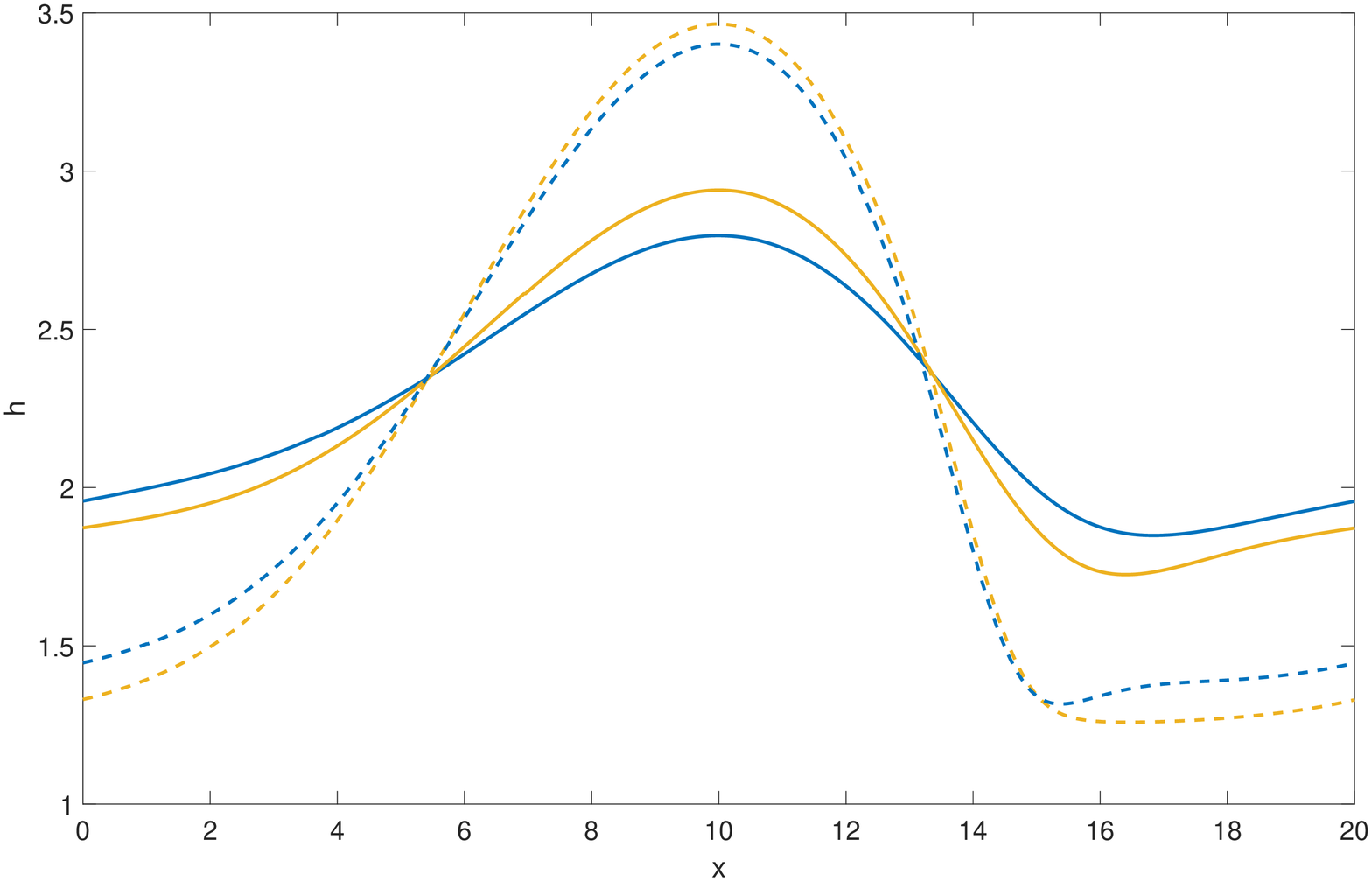}
\includegraphics[height=1.6 in]{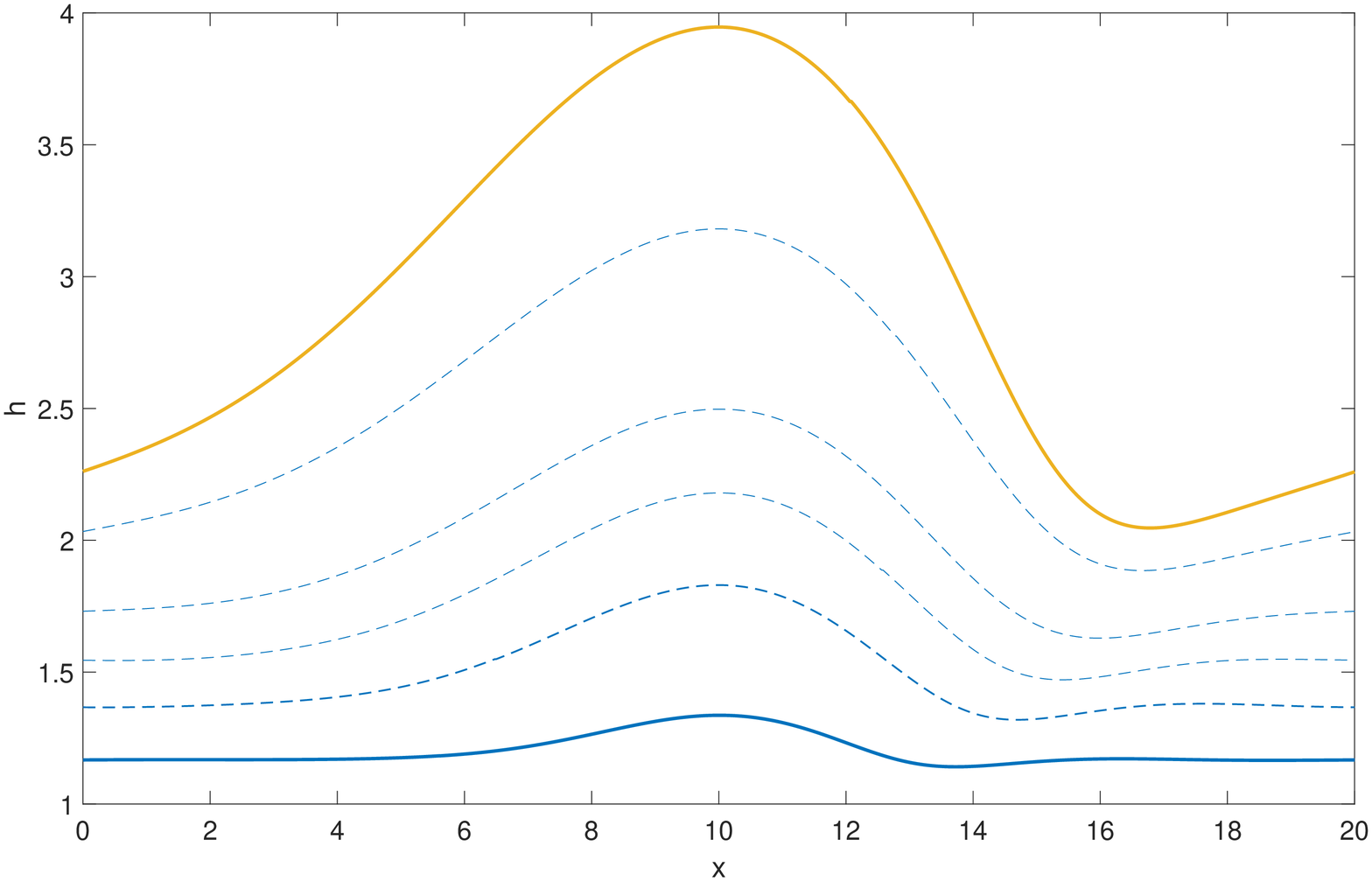}
\caption{\label{two} The left panel illustrates the difference in the maximal height of the traveling waves for the laminar model (solid lines: blue corresponds to $a_1= 0.1$, $b=11$ and yellow corresponds to $a_1=1.5$, $b=13$) and for the plug flow model (dashed lines: blue corresponds to $a=0.2$, $b=10$, $c = 1$ and yellow corresponds to $a=0.4$, $b=12$, $c=3$) using $h_0 = 2.29$ in both models. The right panel shows the variation of the shape of the traveling wave for different $h_0$ values as it ranges from $1.2$ to $3.0$ for fixed values of $a_1 = 1$ and $b = 11$ in the laminar flow model. The non-dimensional velocity of the traveling wave increases from $0.17$ to $2.23$ as $h_0$ increases.}
\end{figure}

\begin{figure}
\includegraphics[height=1.6 in]{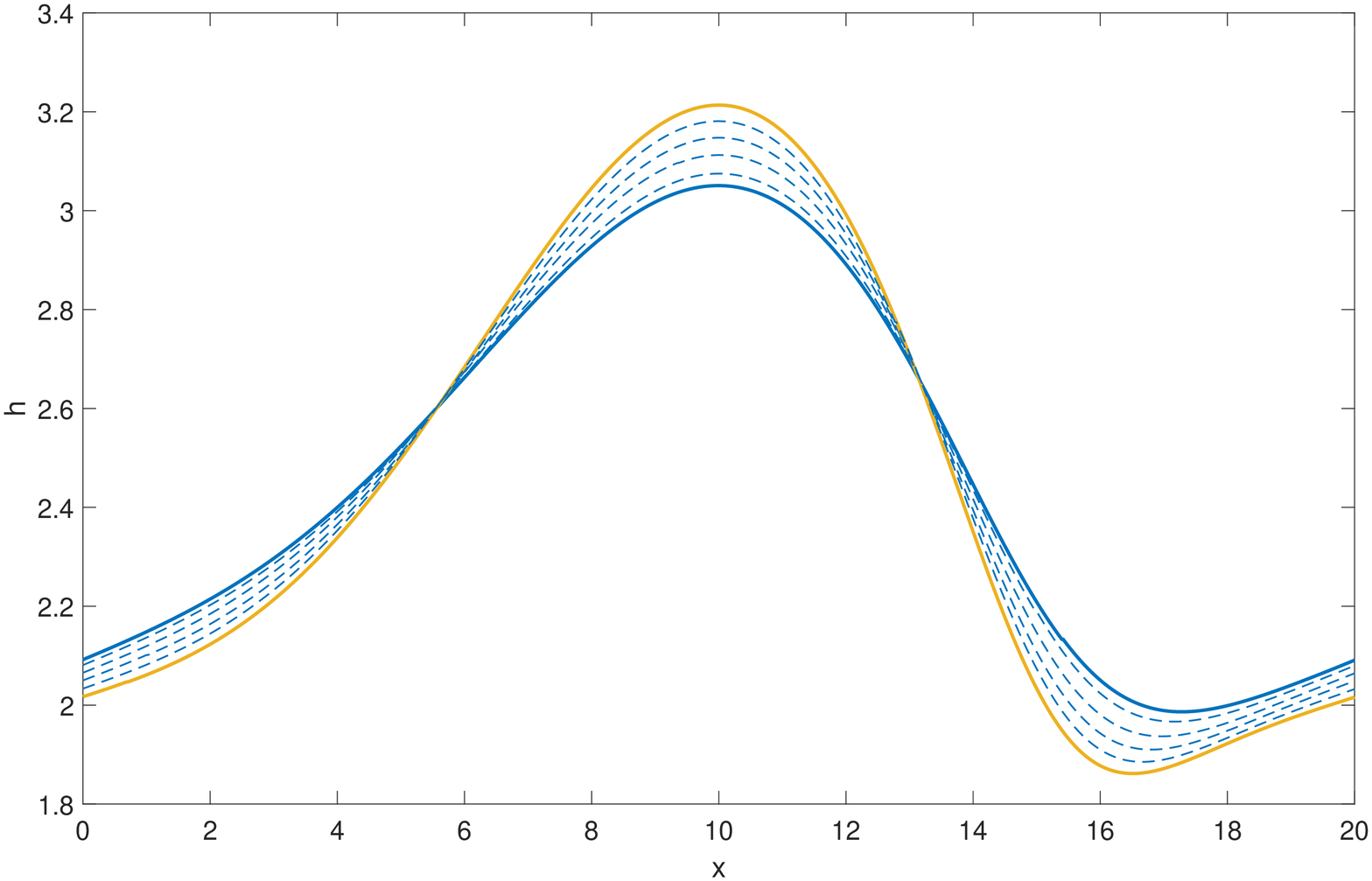}
\includegraphics[height=1.6 in]{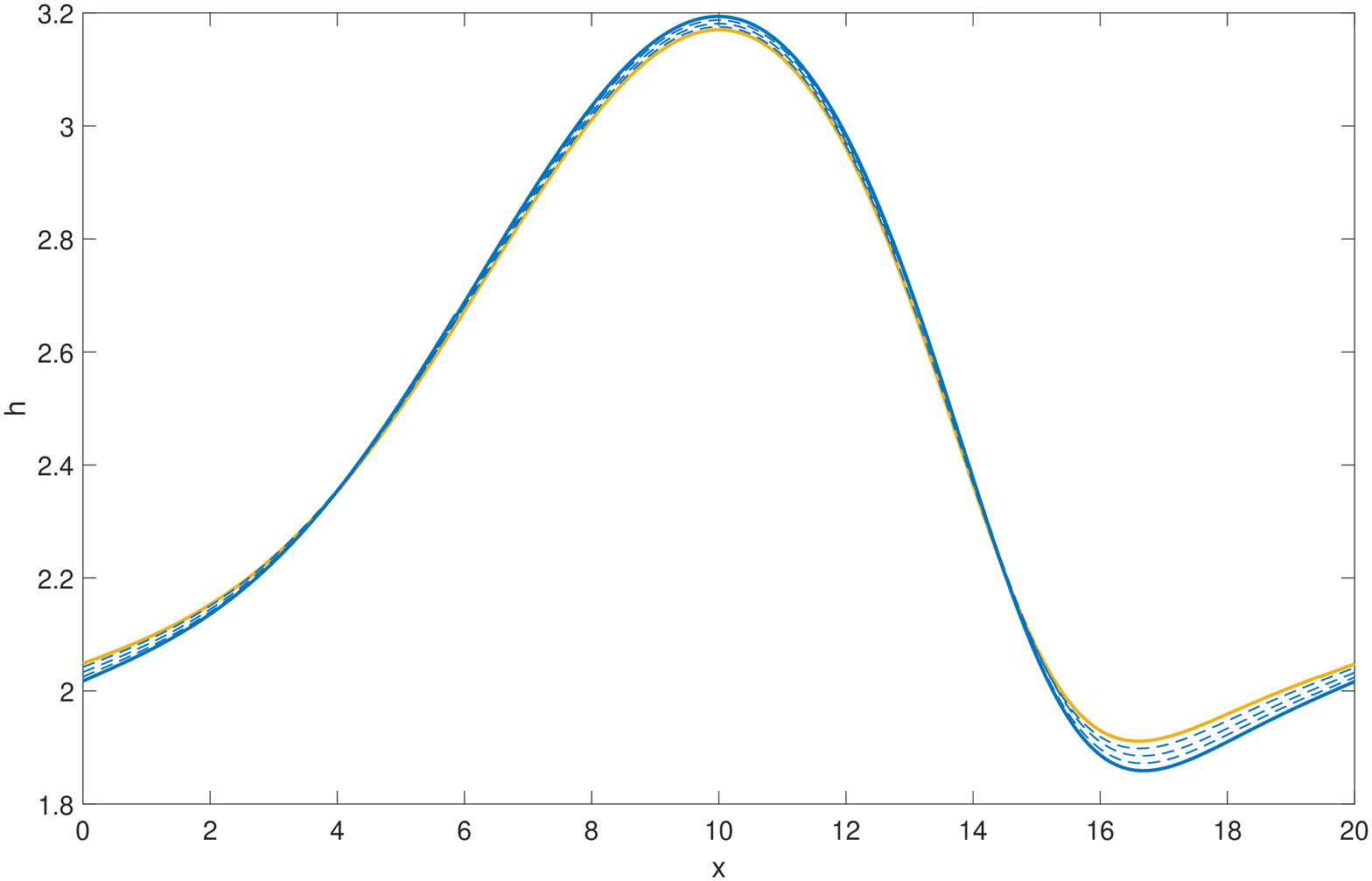}
\caption{\label{dep}  The left panel shows the dependence of the shape of the traveling wave on the parameter $a_1$ as we increase it from $0.1$ (blue) to $1.25$ (yellow) for fixed values $h0 = 2.5$ and $b = 11$ in the laminar model with the non-dimensional velocity of the traveling wave decreasing from $1.338$ to $1.331$.  The right panel shows the dependence of the shape of the traveling wave on the parameter $b$ as we change it from $10$ to $13$ for fixed values $a_1 = 1$ and $h_0 = 2.5$ for the laminar flow model; the non-dimensional velocity of the traveling wave decreases from $1.34$ to $1.32$ in this case.}
\end{figure}

Figures~\ref{two} and \ref{dep} display families of steady traveling wave profiles obtained after longtime simulations of our model equations. On the left panel of Figure~\ref{two} we show the height profile for the laminar flow model (solid lines) for two sets of parameter values, as well as for the plug flow model (dashed lines) also for two different sets of parameter values. We see that the plug flow model leads to a more pronounced peak as compared to the laminar flow model. On the right panel of the same figure, we vary the parameter $h_0$ while keeping the other ones fixed in the laminar flow model. As $h_0$ increases, not only does one obtain a more prominent peak, the speed of the resulting traveling wave also increases substantially. In Section~\ref{sec:ML} we explain how to obtain the speed of the traveling wave from a single snapshot of the steady height and velocity profiles. Parameters $a_1$ and $b$ also affect the shape of the traveling wave, but not as significantly as the film radius $h_0$. As seen in Figure~\ref{dep}, varying $a_1$ (left panel) or $b$ (right panel) does impact the shape of the height profile and the speed of the traveling wave, but much less so than parameter $h_0$.

\section{Experimental Results}
\label{sec:exp}

\subsection{Experiment setup}
\begin{figure}
\begin{center}
\includegraphics[height=2.8 in]{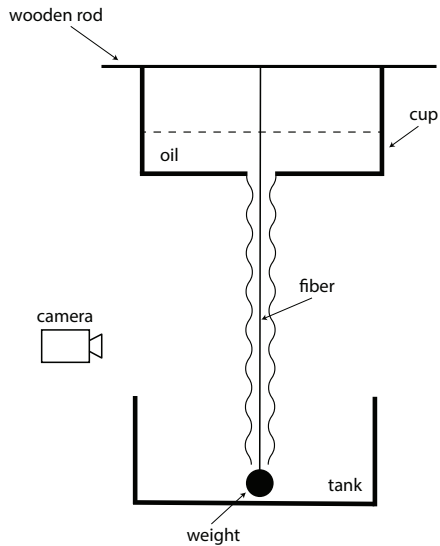}
\caption{\label{setup} Experimental setup scheme.}
\end{center}
\end{figure}

\begin{figure}
\begin{tabular}{c c c c c c}
\begin{tabular}{c}
  \includegraphics[height=2.8in]{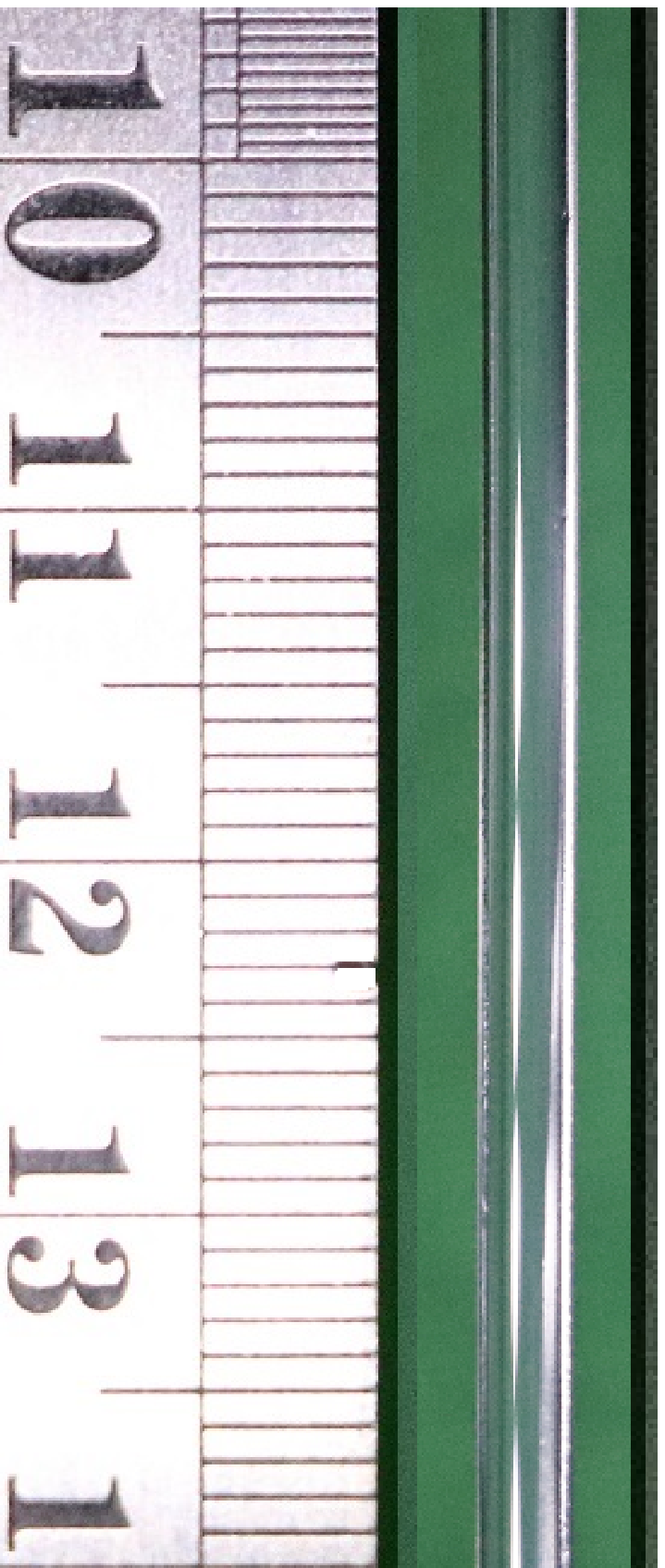} \\
  a) \\
  \end{tabular}
&
\begin{tabular}{c}
  \includegraphics[height=2.8in]{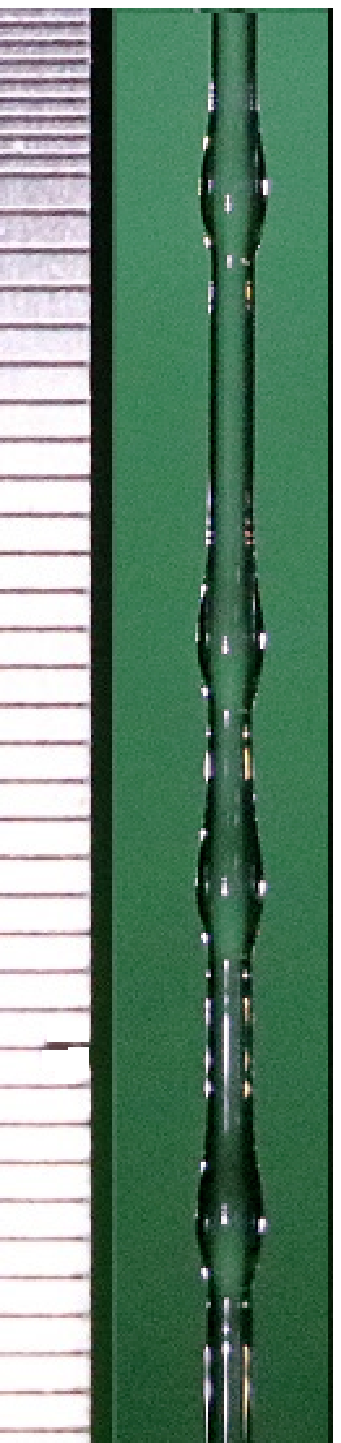} \\
  b) \\
  \end{tabular}
&
\begin{tabular}{c}
  \includegraphics[height=2.8in]{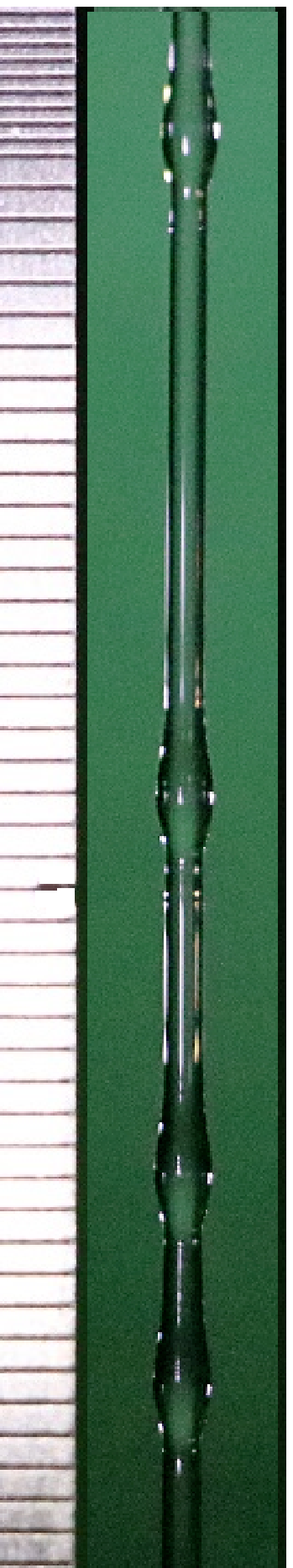} \\
  c) \\
  \end{tabular}
&
\begin{tabular}{c}
  \includegraphics[height=2.8in]{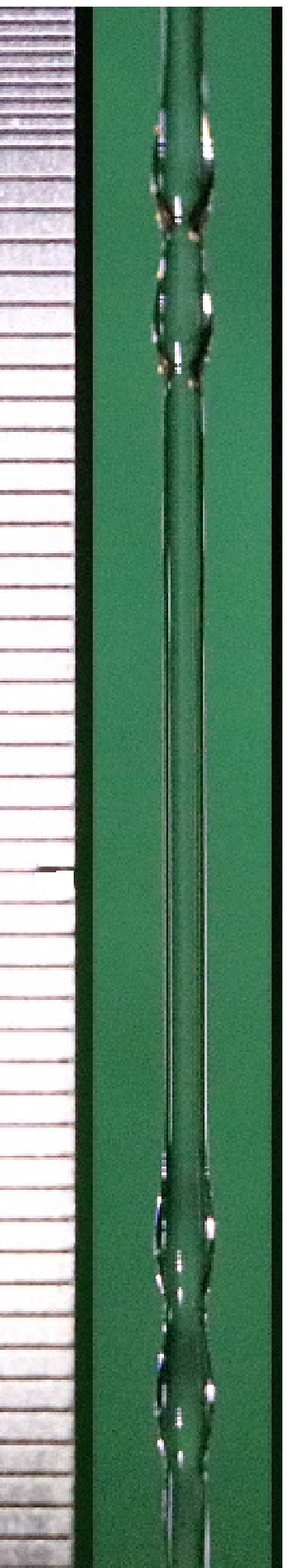} \\
  d) \\
  \end{tabular}
&
\begin{tabular}{c}
  \includegraphics[height=2.8in]{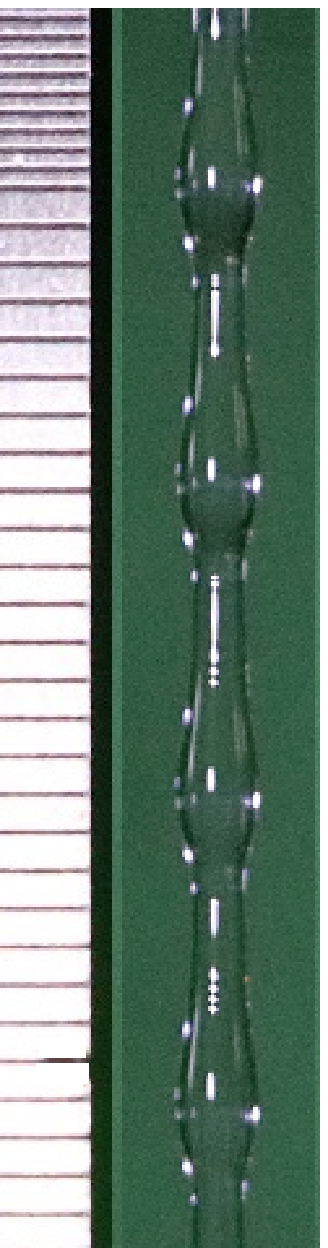} \\
  e) \\
  \end{tabular}
&
\begin{tabular}{c}
  \includegraphics[height=2.8in]{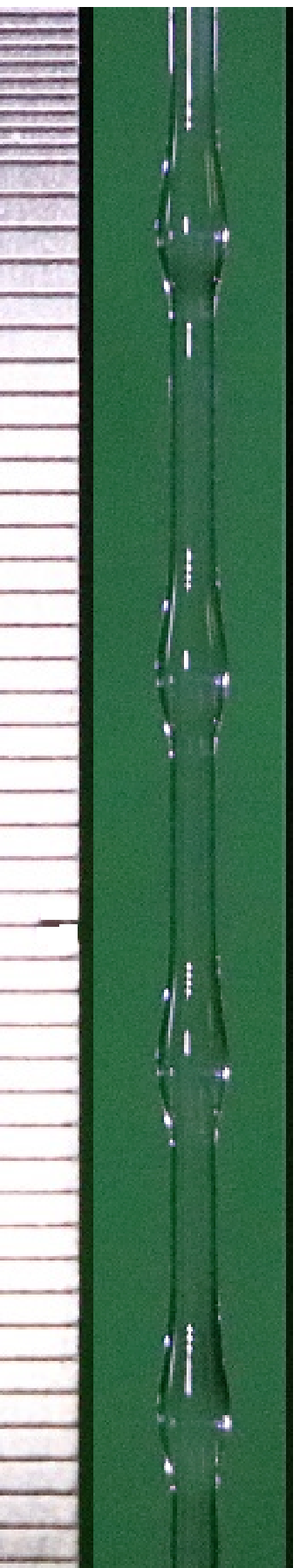} \\
  f) \\
\end{tabular}
 \\
 \end{tabular}
\caption{ \label{drops} a) nearly uniform coating flow over a 2-mm diameter fishing line with a 2.4-mm hole;  b) and c)  non-uniformly spaced droplets on a 1-mm diameter fishing line with the diameter of the hole being 1.6 mm; d) two sets of doublets on a 1-mm diameter fishing line with a hole diameter of 1.6 mm; e) uniformly spaced four-droplet train on a 1-mm diameter fishing line with a hole diameter of 2 mm; f) uniformly distributed train of four droplets on a 1-mm diameter fishing line with  the hole diameter being 2.4 mm (the reference ruler on the left shows 1 mm tick marks, with 1/2 mm ones at the very top).}
\end{figure}

The experimental setup is shown below in Figure~\ref{setup}. Our working fluid is safflower oil with parameter values: density $\rho = 0.928$ g/cm$^3$, absolute viscosity $\mu = 0.0654$ Pa$\cdot$s, surface tension $\sigma = 0.025$ N/m; viscosity was extrapolated to our working temperature of 13$^\circ$C starting with the data from \citet{diam}; density and surface tension data were found online at {\tt https://cameochemicals.noaa.gov/chris/OSF.pdf}. Safflower oil was placed in a cup with a hole drilled in the center. A nylon fishing line was passed through the hole and tied to a hanging weight at the bottom in order to maintain a vertical line. The line was threaded through a wooden rod that was placed horizontally across the top of the cup to hold the fiber in place. Our setup was modeled after those presented in other papers examining droplet flow on a vertical fiber, e.g., \citet{KDB} and \citet{craster2006viscous}. One main difference in our setup is the use of larger diameter fishing lines on which different regimes of coating flows were observed. Cups with hole sizes ranging from $1.6$ to $2.4$ mm were used in order to control the flow rates, and both 1 and 2 mm diameter fishing lines were used. Snapshots of the flow were photographed with a green screen background using a Canon EOS Rebel T6s at a shutter speed of 1/4000 s. Video was also captured in order to determine droplets velocity.

\subsection{Observations}

We lubricated our fishing lines with the oil before running the experiments since on dry fishing lines we would see a very strong instability near the advancing side of the leading drop, causing tiny droplets to be scattered in all directions while propagating down the line.

For the 2-mm diameter fishing line through the 2.4-mm size hole we observed a nearly uniform coating flow. We could see a very slight waviness of this film near the edge so we were certain that some very small amplitude waves were present; see a) in Figure~\ref{drops}. This almost uniform coating flow behavior was not mentioned in prior experimental publications such as \citet{KDB} and \citet{craster2006viscous}, most likely due to the thinner lines they used. For the 1-mm diameter fishing line, with hole diameters ranging from 1.6 to 2.4 mm, different distributions of droplets flowing down the line were observed. For the 1.6-mm diameter hole with the 1-mm diameter fishing line we observed non-evenly spaced trains of droplets: see b) and c) in Figure~\ref{drops}, and we also obtained interesting doublet configurations: e.g., d) in Figure \ref{drops}. The reason the distances are so different in this regime is that droplets interact with each other as they do not have the same size and speed.  For the 2-mm hole we observed uniformly spaced droplets, see e) in Figure \ref{drops}, and for the 2.4-mm hole we also observed uniformly spaced droplets that were bigger than the previous case with slightly longer periods: see f) in Figure \ref{drops}. The distributions of droplets we observed on the 1-mm line are qualitatively similar to ones reported in \citet{KDB} and \citet{craster2006viscous}: in Figure~\ref{drops},  b), c) and d)  correspond to the Convective Instability Regime, and e) and f) correspond to the Rayleigh-Plateau and Isolated Droplets Regimes. The table below lists the corresponding parameter values measured at temperatures of about 13--15$^\circ$C.

\begin{center}
      \begin{tabular}{||c|c|c|c|c|c||}
        \hline
        \small{Drill Size} & \small{Hole Size} & \small{Flow Rate} & \small{Droplet Size} & \small{Gap Size} & \small{Droplet Velocity}\\
         & \small{(mm)} & \small{(g/s)} & \small{(mm)} & \small{(mm)} & \small{(mm/s)}\\
         \hline
        1/16 & 1.6 & 0.0067 & 3-5 & various & 10 \\
        5/64 & 2 & 0.0366 & 4 & 3 & 15 \\
        3/32 & 2.4 & 0.0833 & 5 & 7 & 30 \\
        \hline
      \end{tabular}
\end{center}

In all experiments we noticed that a steady ``ring'' of fluid always developed around the cup exit hole and new droplets always detached from this ring without changing the shape of the ring.

%

\section{Application of Machine Learning to Model Validation}
\label{sec:ML}

Using the approximate physical properties of the oil mentioned earlier ($\rho=928$ kg/m$^3$, $\mu=0.0654$ Pa$\cdot$s and $\sigma=0.025$ N/m), with $g=9.8$ m/s$^2$ and for the 1-mm diameter fiber whose radius is $R=0.5$ mm, for a nominal film thickness $H-R=0.75$ mm (estimated visually), the velocity scale $U_1=2 g R^2/\nu$ ends up being 6.95 cm/s and the dimensionless parameters for the laminar flow model turn out to be $a_1=0.987$, $b=11.0$ and $h_0=2.5$. Also, if we define the Reynolds number as Re=$U_1 R/\nu$ we get Re$=0.493$, which is low enough that the laminar flow model should be reasonable. However, since the physical properties are approximate and the film thickness is not exactly measured, we apply Machine Learning to find the closest set of parameters that match the model predictions to the experimental observations.

Using COMSOL we created a labeled set of 182 data files in which a final snapshot of the steady traveling wave profile $h(x)$ and mean velocity $u(x)$ were saved as a function of discretized $x$ at 961 equally spaced nodal points including the end points on the interval $x \in [0,L]$ with $L = 20$. In those data sets, random values of parameters $a_1$, $b$ and $h_0$ were chosen in the neighborhood of the original estimates. Note that parameter $c_1$ is fixed at $c_1=4$ in the laminar flow model. We took uniformly distributed random values of $h_0 \in [2.2, 2.8]$, of $a_1 \in [0.1, 1.1]$, and of $b \in [10, 13]$. In the simulations, we started with the initial condition $h(x,0)=h_0+0.1 \sin(2 \pi x/L)$ and integrated to a long enough time that a steady traveling wave profile was reached. The label of each data file included values of the random parameters that generated it. While all three parameters varied at random in these simulations, the set of profiles obtained were roughly similar to those displayed in Figures~\ref{two} and \ref{dep} in which two of the parameters were held fixed while the third varied. 

We then trained our learning algorithm on this data set so that we would be able to ``predict'' the values of the dimensionless parameters if we were given a certain discretized profile $h(x)$. That way, we could discretize the experimentally observed profile and find out the set of parameter values that would generate that profile. For the supervised learning we used the so-called Gradient Boosting Regressor (GBR) based on an ensemble of decision trees. We applied the Holdout Method for cross validation by taking $20\%$ of our data as a test set.

\begin{figure}
\begin{tabular}{l l}
a) \includegraphics[height=1.4 in, keepaspectratio]{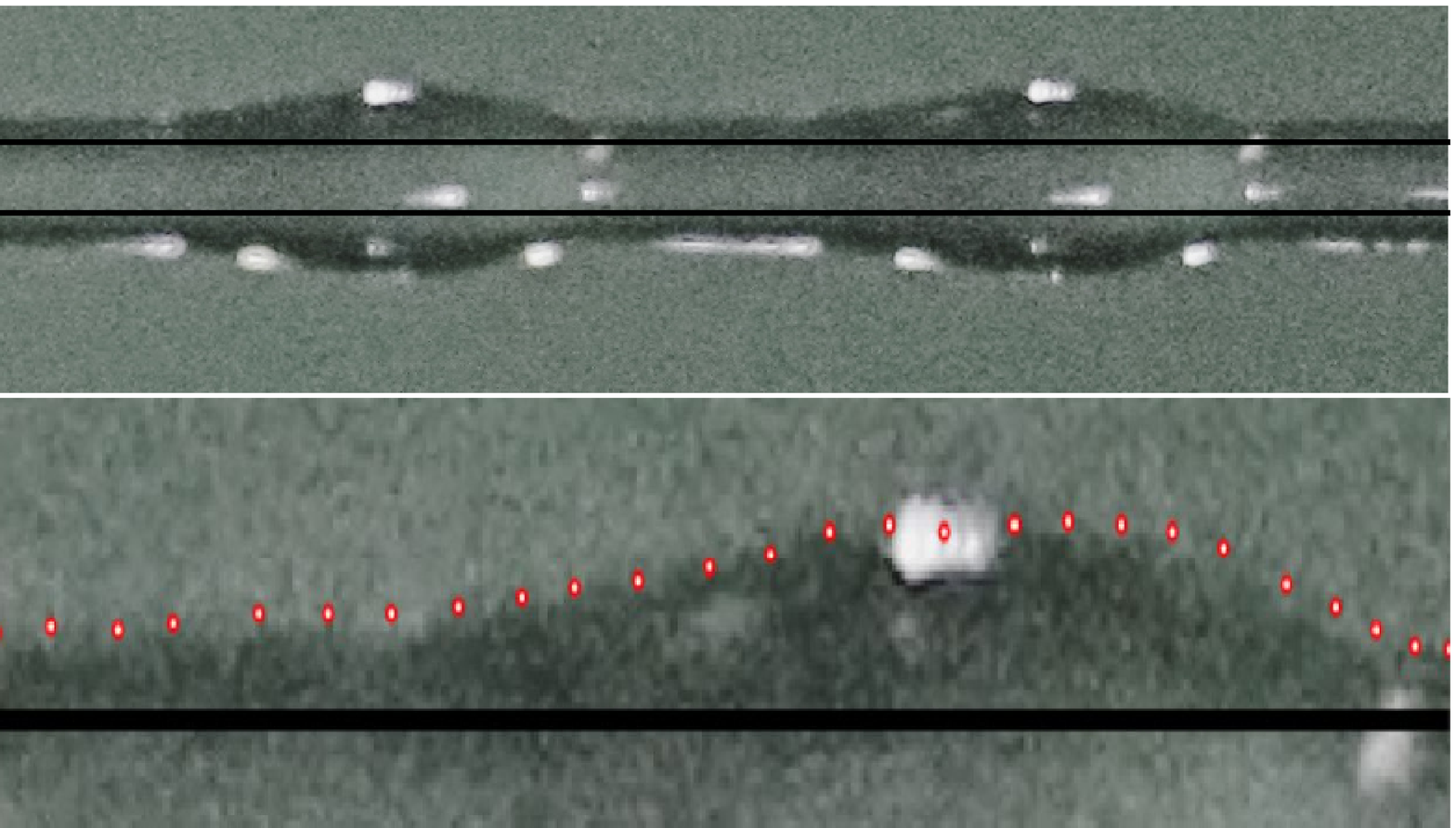} & b) \includegraphics[height=1.7 in, keepaspectratio]{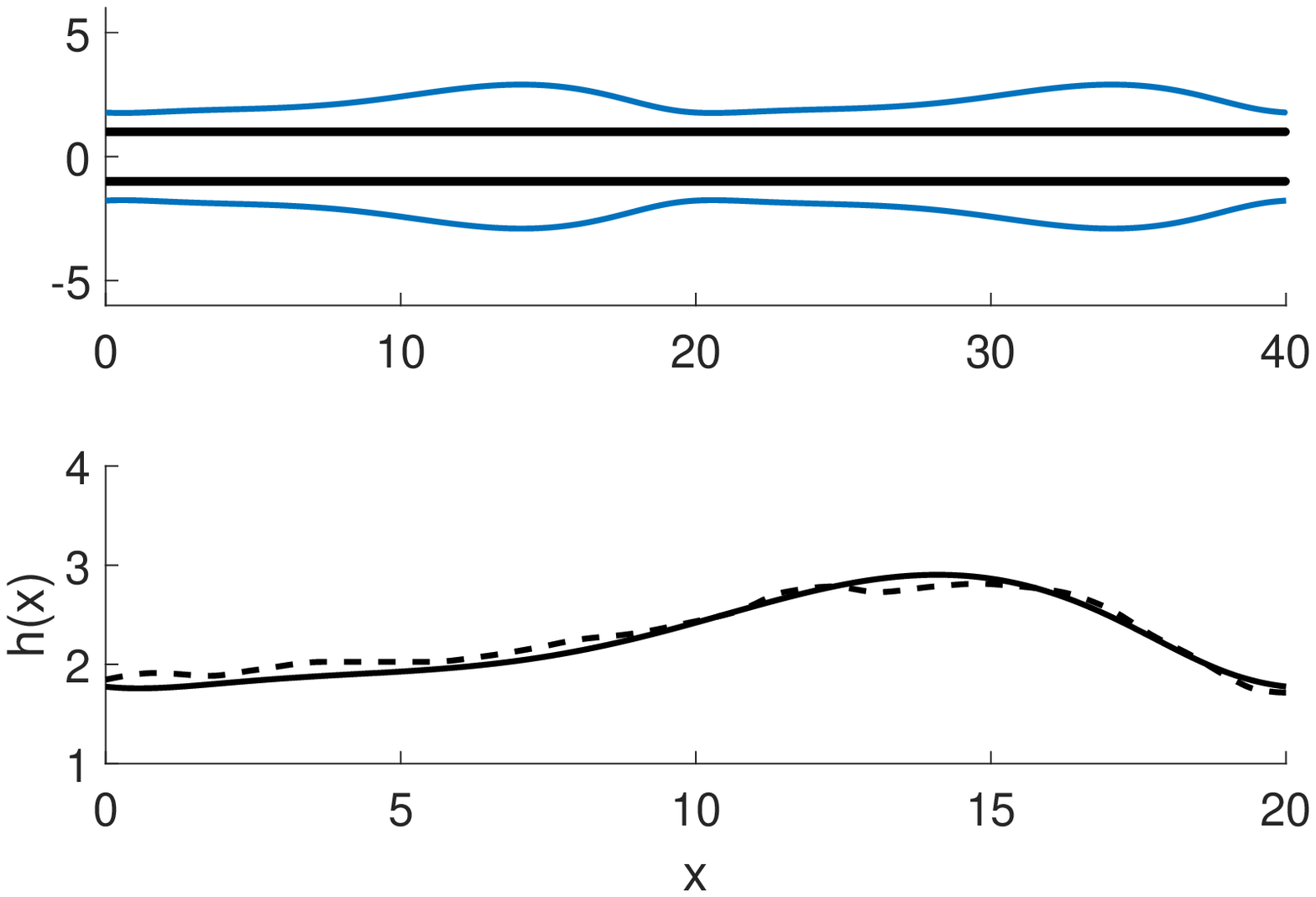}  \\
\end{tabular}
\caption{\label{edge} a) Zoomed photo of the droplets with edge detection indicated below by the series of red dots. b) Match between the edge of the drop and a numerical simulation corresponding to the parameters found by Machine Learning: $h_0 = 2.29$,  $b = 12.32$  and  $a = 0.95$.}
\end{figure}

To compare our numerical simulation result (steady traveling wave $h$ function) with an experimental shape of a droplet we zoomed in on the droplet photo and traced the edge by manually inserting marker points using \emph{Mathematica} (red points in Figure \ref{edge}) to discretize the profile. We then used interpolation of the captured coordinates of the marker points to represent this profile in our numerical 961-point data format and used these experimental data as an input for our trained GBR model to obtain the following predictions for the parameter values: $a=0.95$ and $b = 12.32$. The value of $h_0$ could be obtained readily through a conservation of volume constraint. Namely, since
$$ \int_0^{L} (h_0 + 0.1 \, \sin{(2 \pi x/L)})^2 \, dx = \int_0^{L} h(x,t)^2 dx\,, $$
using the experimental profile on the right-hand side yielded $h_0=2.29$. We used the initial perturbation $\sin{2\pi x/L}$ to speed up the convergence of simulations to a steady traveling wave on the computational domain $ x \in [0, L]$ with $L=20$ as the wavenumber $k=L/(2\pi)$ is close to the most unstable mode in the linearized about $h_0$ model. We compare the experimental results and the numerical simulations by moving the maximum height for both to the same point of the grid. The bottom right panel in Figure~\ref{edge} displays the two curves and the good agreement between them.

\begin{figure}
\begin{tabular}{c c}
  a) \includegraphics[height=1.4 in]{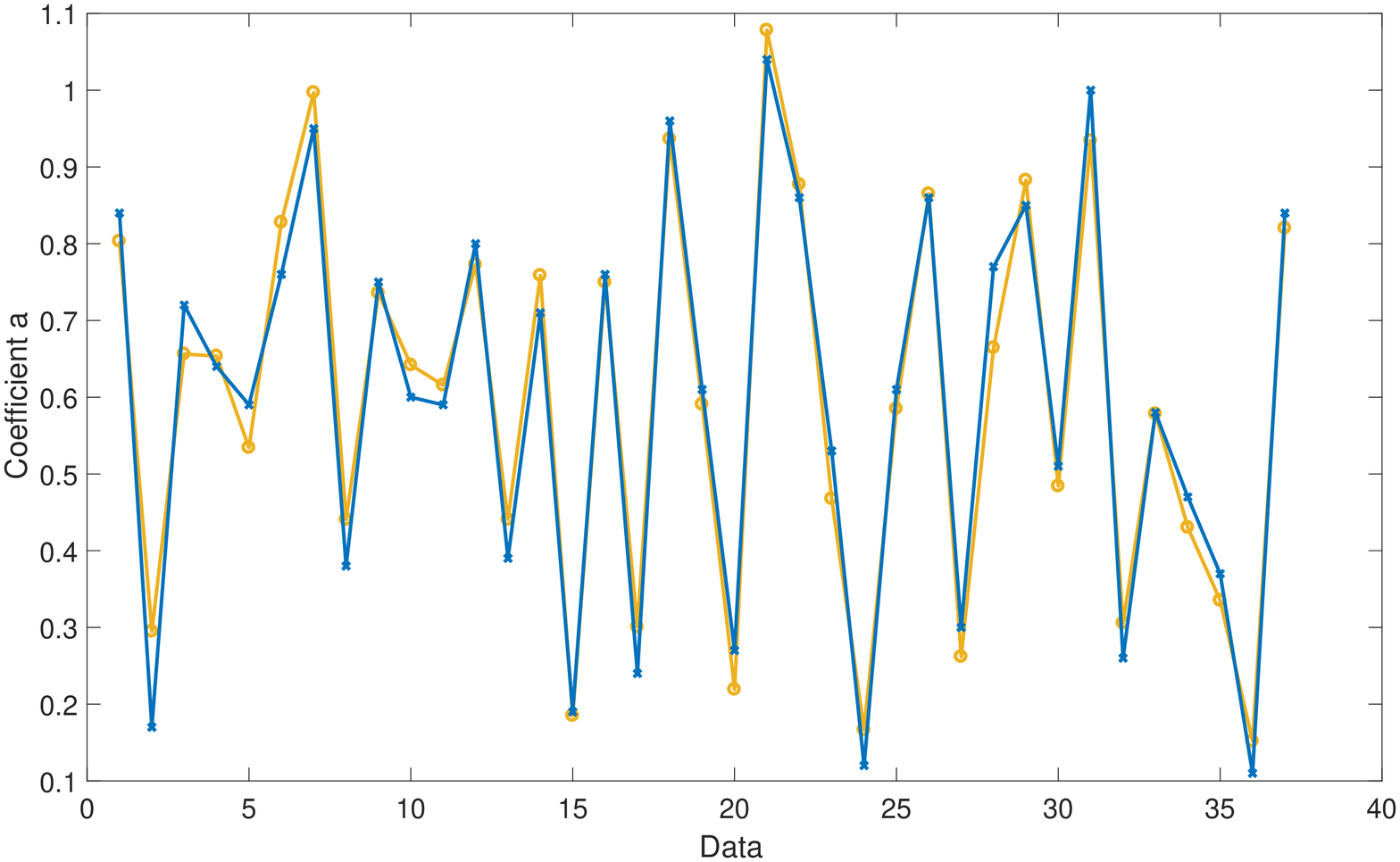} & d) \includegraphics[height=1.4 in]{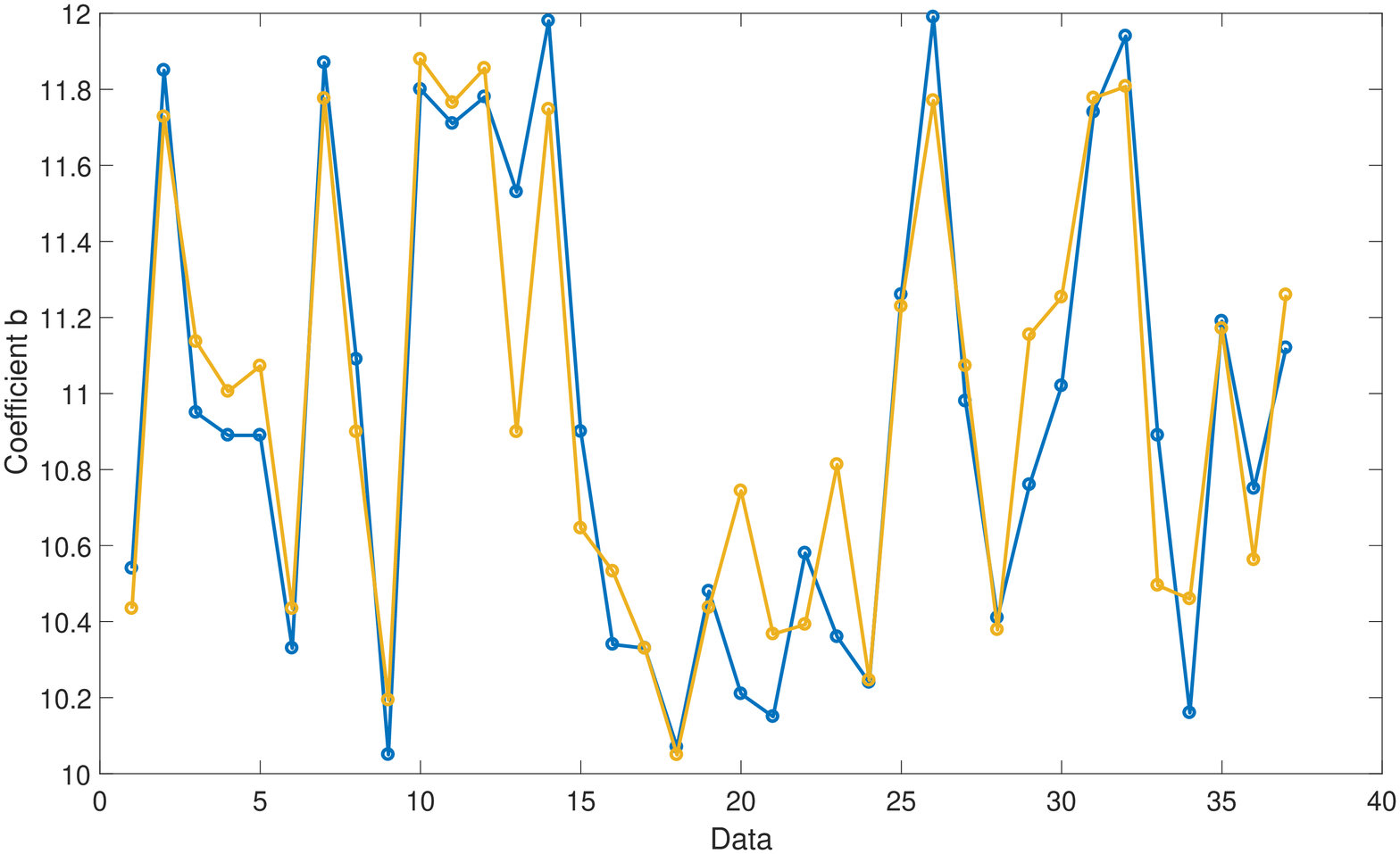} \\
  b) \includegraphics[height=1.4 in]{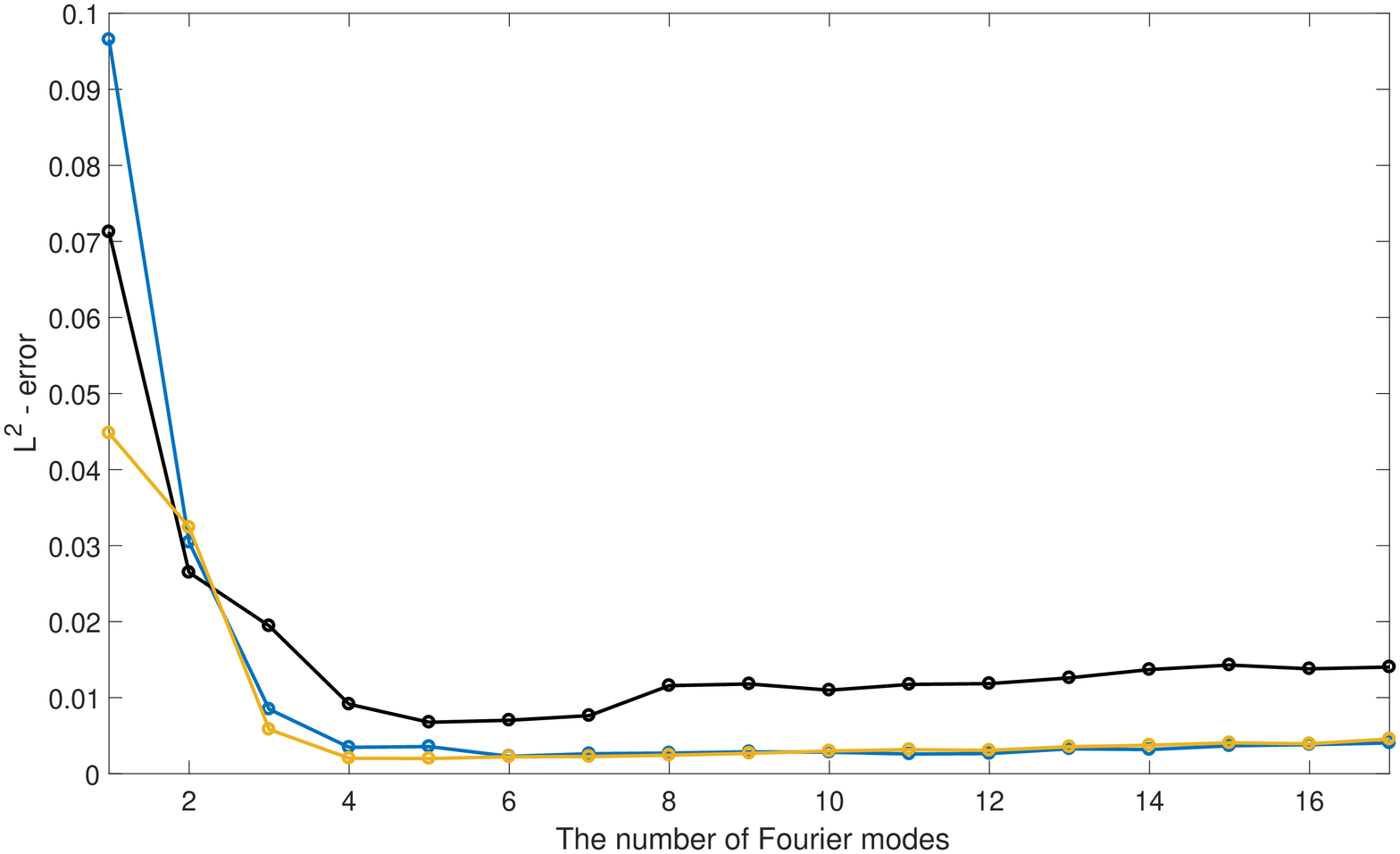} & e) \includegraphics[height=1.4 in]{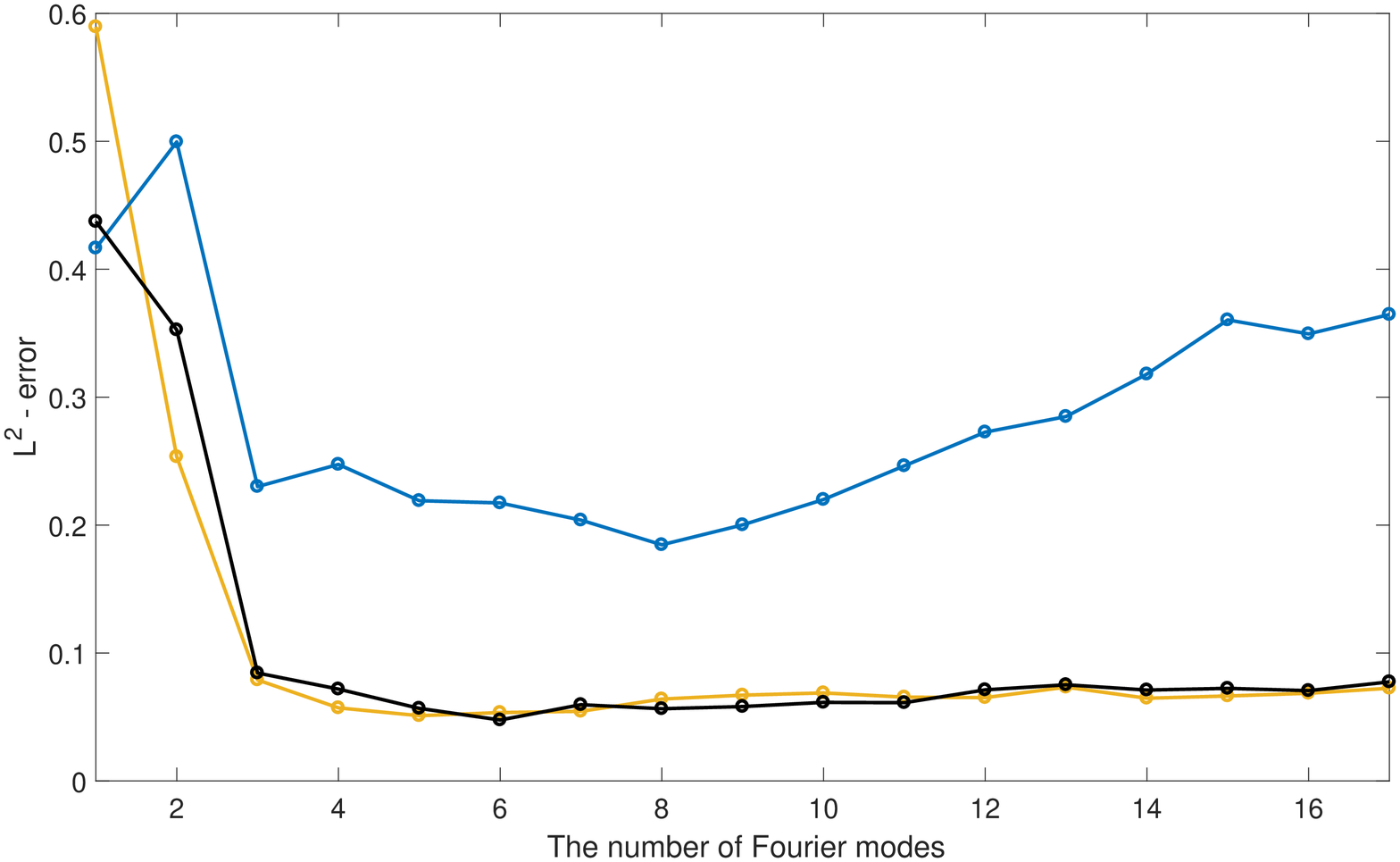} \\
  c) \includegraphics[height=1.4 in]{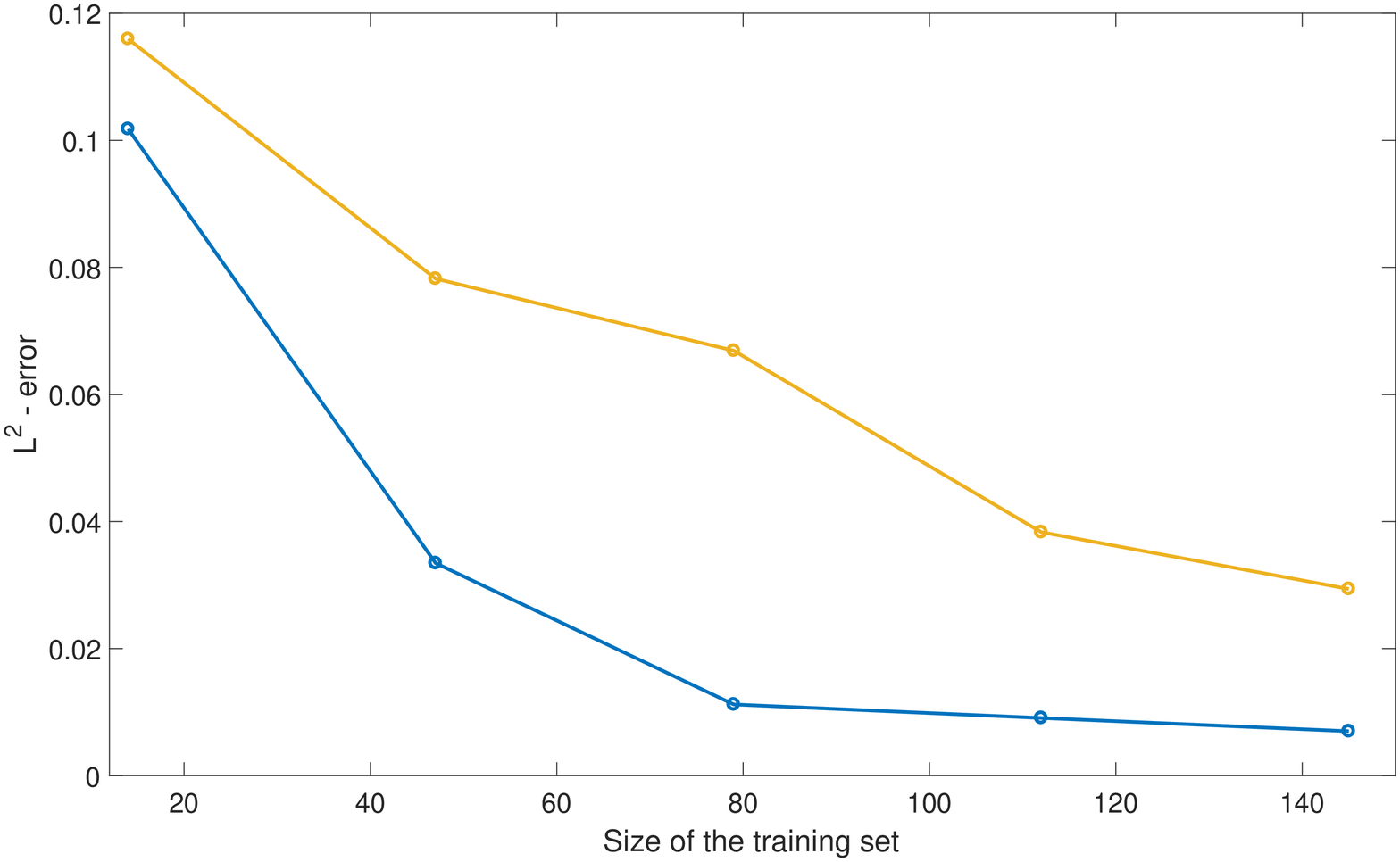} & f) \includegraphics[height=1.4 in]{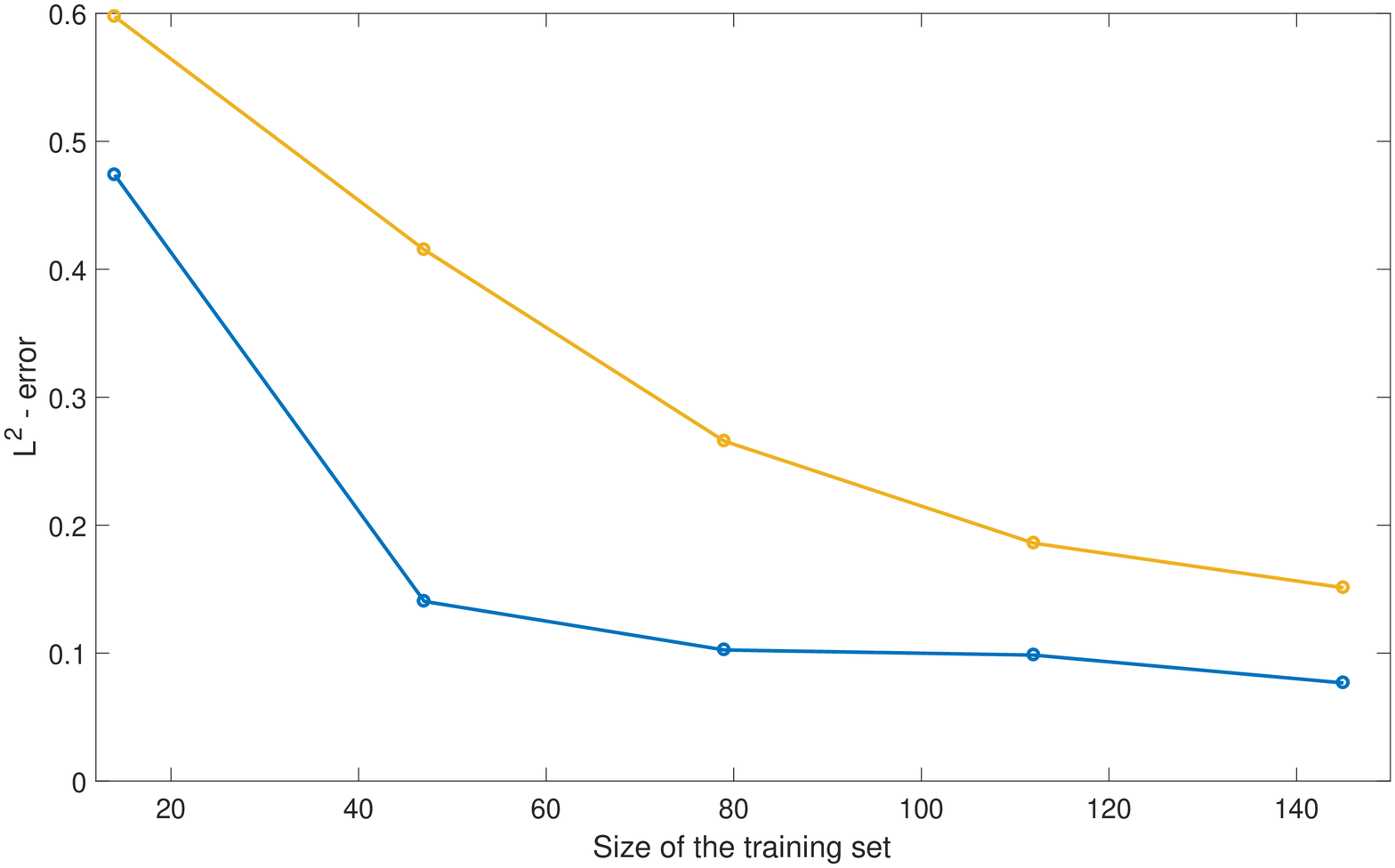} \\
 \end{tabular}
 \caption{\label{ml} \label{gbr} a) Plot of actual (blue) vs predicted (yellow) values of $a_1$ found by the GBR, b) Accuracy of the parameter $a_1$ value on a test set vs the number of first Fourier coefficients used (blue: $h$, black: $1/(h^2 -1)$, yellow: both), c) Learning curve: the accuracy of the parameter $a_1$ vs the size of the training dataset (yellow: 961 grid values, blue: first 6 Fourier coefficients), d) plot of actual (blue) vs predicted (yellow) values of $b$ found by the GBR, e) Accuracy of parameter $b$ on a test set vs the number of first Fourier coefficients used (blue: $h$, black: $1/(h^2 -1)$, yellow: both), f) Learning curve: the accuracy of the parameter $b$ vs the size of the training dataset (yellow: 961 grid values, blue: first 6 Fourier coefficients). }
\end{figure}

In the process of training the GBR (see Figure \ref{gbr}) we discovered that different tailored data sets could be used to improve the accuracy for the prediction of coefficients $a_1$ and $b$. The optimal approach for obtaining the coefficient $a_1$ is to use as training data the first $6$ coefficients of the Fourier series of the periodic traveling wave $h(x)$, and for the coefficient $b$ to use as training data the first $6$ coefficients of the Fourier series of the function $1 /(h(x)^2 -1)$. This difference was a good indication that the shape of the traveling wave $h(x)$ is mostly defined by the value of the coefficient $a_1$ and the mean velocity $u(x)$ is mostly governed by the value of the coefficient $b$. The derivation below explains the relation between the speed $u(x)$ and the reciprocal of the area $1 /(h(x)^2 -1)$.

\begin{figure}
\includegraphics[height=1.7 in]{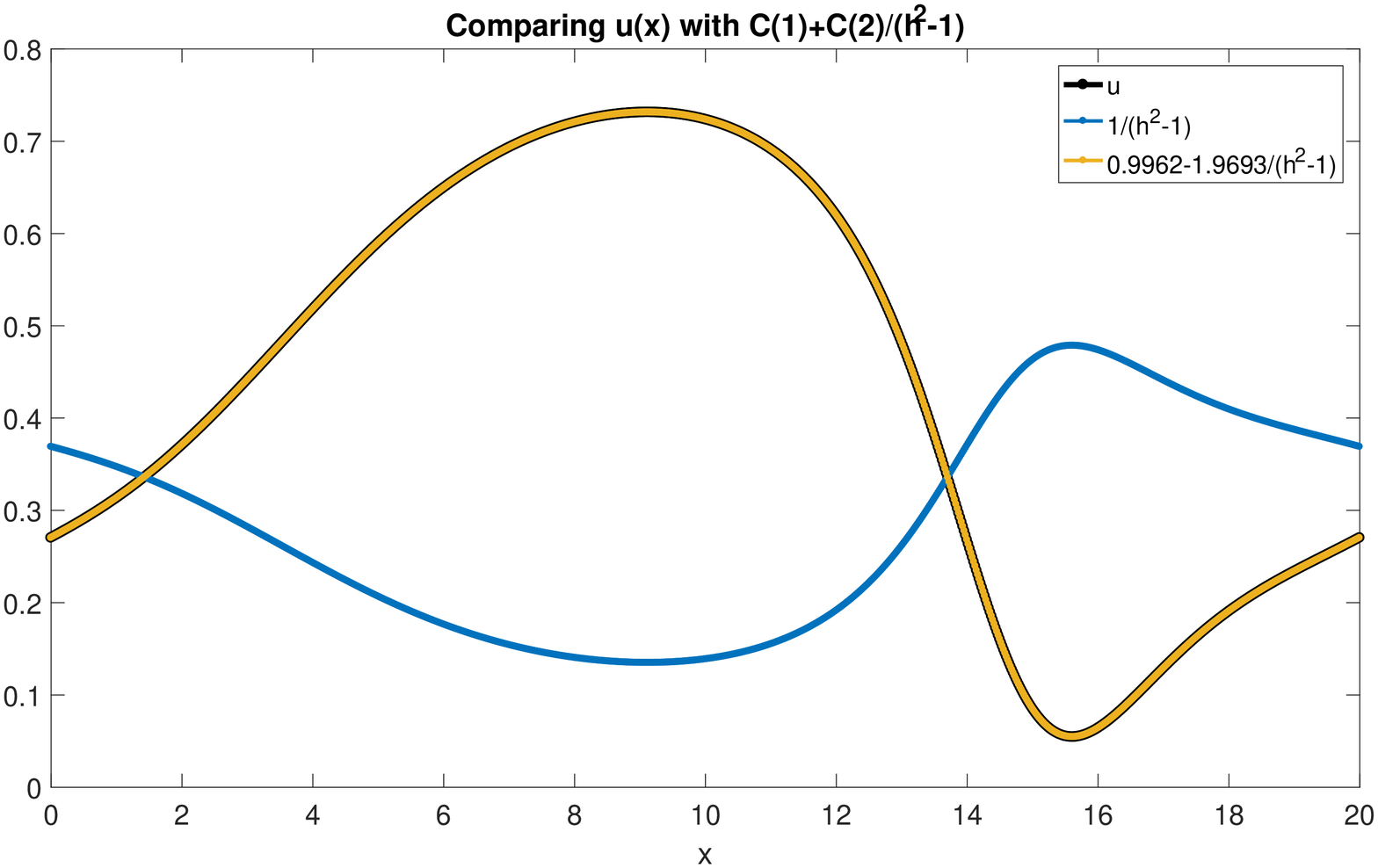}
\includegraphics[height=1.7 in]{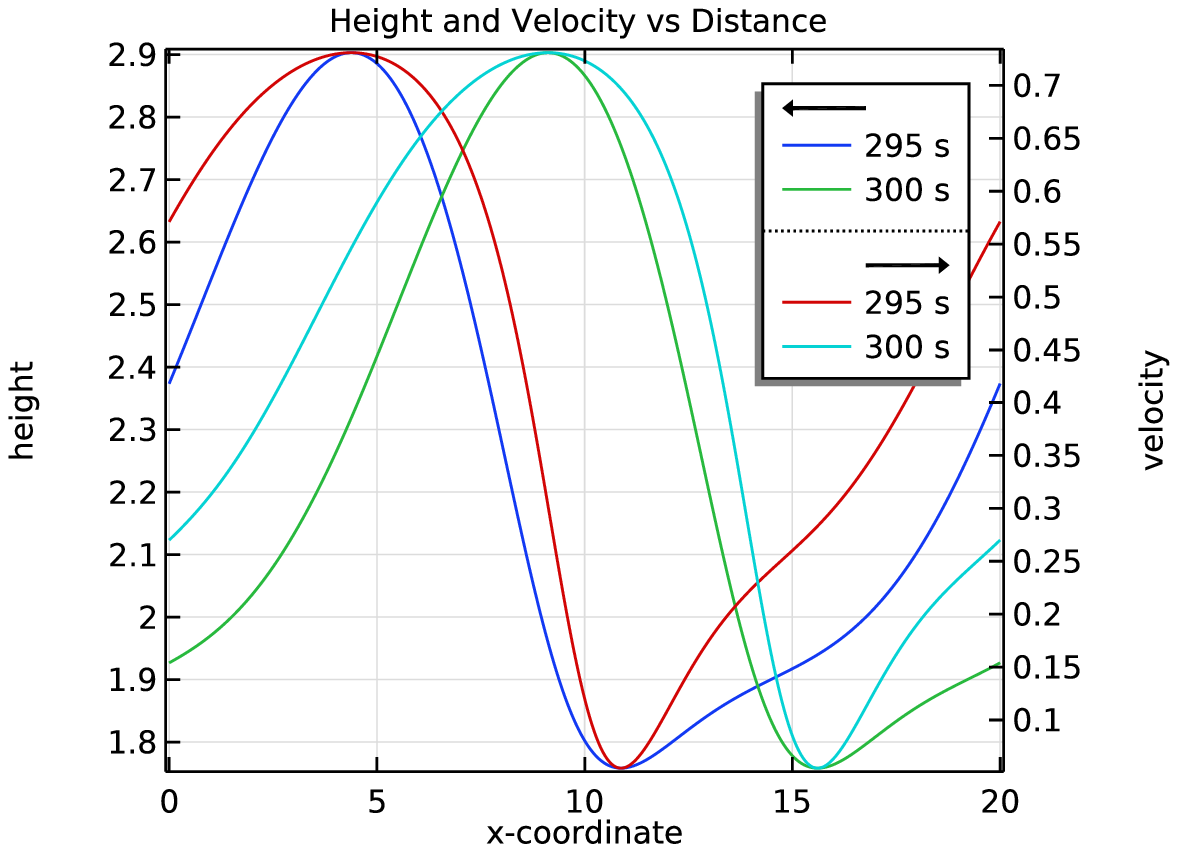}
\caption{\label{fitsp} The plots on the left illustrate the perfect fit obtained when $u(x)$ is fit to a function of the form $C_1+C_2/(h(x)^2-1)$. The blue curve shows $1/(h^2-1)$ while the other curve shows $u(x)$ and its best fit that appear superposed. The plots on the right show the height and velocity profiles at two different times and confirm that the traveling wave speed $V$ is matched with high accuracy.}
\end{figure}

Using the traveling wave ansatz that $h(x,t)$ and $u(x,t)$ only depend upon the dimensionless traveling wave coordinate $z = x - Vt$, with $u=u(z)$ and $h=h(z)$, the conservation of mass equation (\ref{eq:mass-nondim-laminar}) results in the ordinary differential equation:
$$ -V(h^2)' + a_1 (u(h^2 -1))' = 0$$   
whose solution after some algebraic manipulation provides
$$u(z) = \frac{V}{a_1} + \frac{C}{h(z)^2 -1}.$$
Here, primes denote derivatives with respect to $z$. So $u(z)$ does depend directly on the ${1}/{(h(z)^2 -1)}$. This computations also suggest a simple method of finding the traveling wave velocity $V$ from just one snapshot of the profiles $h(x)$ and $u(x)$ at some late time. One can find the dimensionless traveling wave speed $V$ by just fitting  the profile of $u(x)$ to a function of the form $u(x) = C_1 + C_2 \, \frac{1}{h(x)^2 -1}$ and using the best fit coefficient $C_1$ to define $V = a_1 C_1$, see Figure \ref{fitsp}. To obtain the dimensional traveling wave speed from the coefficient $C_1$, we multiply the latter by velocity scale $U_1$. 


\section{Discussion}
\label{sec:disc}

One of the main modeling challenges related to droplets sliding down a fiber is the mismatch between the experimentally observed velocity of the drops and that obtained from numerical simulations based on a model. In the models by \citet{KDB} and \citet{craster2006viscous} this mismatch of velocities was close to 40\%. In \citet{Bert2019} an artificial stabilizing term was introduced to improve the match between the traveling wave droplet velocity and the experiments.  Our laminar flow model provides  closer values of the traveling wave velocity to the experiments for smaller values of $h_0$. However, for the value $h_0=2.29$ obtained above, the predicted velocity is about 6 cm/s, whereas the experimentally observed droplet velocity seemed to be about half of that. The other challenging part in modeling is to obtain clear qualitative transition criteria between the various regimes observed in experiments. Another still unsolved puzzle is the influence of the size or geometry of the hole (i.e., the source) on the distribution of droplets and their dynamics while flow rate of the fluid is kept constant. 


While in this paper we focused most of our attention on the laminar flow model appropriate for low Reynolds numbers, our plug flow model should apply to well-mixed possibly turbulent flows with a velocity profile closer to plug flow. Just to see whether this is physically feasible, consider a hypothetical system with the following assumed parameters: suppose the fiber radius $R$ is $2$~mm and the film thickness $T=H-R$ is about the same size as the fiber radius. Take the working fluid to be water whose viscosity is much less than the oil. We can approximate the corresponding Reynolds number for a steady state flow of this type. The mean wall shear stress $\tau$ is expressed in terms of the Darcy–Weisbach friction factor $f_{D}$ and average fluid velocity $U$ as
\[
    \tau = \frac{1}{8} f_{D} \, \rho  U^2
\]
The force balance between the drag force from the wall and gravity gives us
\[
    2 \pi R \tau = \rho g \pi ( (R + T)^2 - R^2 )
\]
which can be solved to obtain $\tau = 78.4$~Pa (we take the density of water to be $\rho=1000$~kg/m$^3$ and its viscosity to be $\mu=0.001$~kg/(m~s)). The Colebrook–White correlation for a smooth surface relates the friction factor to the Reynolds number by
\[
    \frac{1}{ \sqrt{f_{D}}} = -2 \log( \frac{2.51}{ \mbox{Re} \sqrt{f_{D}}})
\]
We substitute the expression for $f_{D}$ in terms of $\tau$ and $U$, and $\mbox{Re} = \frac{\rho U T}{ \mu } $ and obtain a transcendental  equation for $U$, whose solution yields the mean velocity and corresponding Reynolds number as 
\[
    U = 11.31 \mbox{m/s}, \quad \mbox{Re} = 4.5\times 10^4 \,.
\]
The result shows that under some practical assumptions, the film flow on fiber could be in a turbulent regime. Under the assumptions above, the parameters $a,b,c$ in our model would have values
\[
    a \approx 102, \quad b \approx 1.84,\quad c \approx 0.072\,. 
\]
Exploration of such turbulent regimes and their experimental investigations are left for future work. 

\bibliographystyle{jfm}
\bibliography{YR_AN_LD_MC_2021}

\end{document}